\documentclass[prd,nofootinbib,english]{revtex4}
\usepackage{amsmath}
\usepackage{amsfonts,color}
\usepackage{amsmath}
\usepackage{amsthm}
\usepackage{amsfonts,color,xcolor}
\usepackage{amssymb,float}
\usepackage{accents}
\usepackage{hyperref}
\hypersetup{
    colorlinks=true,
    linkcolor=blue,
    filecolor=magenta,      
   citecolor=blue
}
\usepackage[utf8]{inputenc}
\allowdisplaybreaks

\newcommand{\lc}[1]{\accentset{\circ}{#1}}

\newcommand{\dd}{{\rm d}}

\newtheorem{theorem}{Theorem}
\usepackage{bookmark}

\usepackage{enumitem}

\usepackage{mathtools}

\definecolor{mygreen}{rgb}{0,0.7,0}

\begin{document}
\title{Scalarized Black Holes in Teleparallel Gravity}

\author{Sebastian Bahamonde}
\email{sbahamondebeltran@gmail.com, bahamonde.s.aa@m.titech.ac.jp}
\affiliation{Department of Physics, Tokyo Institute of Technology
1-12-1 Ookayama, Meguro-ku, Tokyo 152-8551, Japan.}
\affiliation{Laboratory of Theoretical Physics, Institute of Physics, University of Tartu, W. Ostwaldi 1, 50411 Tartu, Estonia.}

\author{Ludovic Ducobu}
\email{ludovic.ducobu@umons.ac.be}
\affiliation{Nuclear and Subnuclear Physics, University of Mons, Mons, Belgium}

\author{Christian Pfeifer}
\email{christian.pfeifer@zarm.uni-bremen.de}
\affiliation{ZARM, University of Bremen, 28359 Bremen, Germany}

\begin{abstract}
Black holes play a crucial role in the understanding of the gravitational interaction. Through the direct observation of the shadow of a black hole by the event horizon telescope and the detection of gravitational waves of merging black holes we now start to have direct access to their properties and behaviour, which means the properties and behaviour of gravity. This further raised the demand for models to compare with those observations.
In this respect, an important question regarding black holes properties is to know if they can support ``hairs". While this is famously forbidden in general relativity, in particular for scalar fields, by the so-called no-hair theorems, hairy black holes have been shown to exist in several class of scalar-tensor theories of gravity.
In this article we investigate the existence of scalarized black holes in scalar-torsion theories of gravity.
On one hand, we find exact solutions for certain choices of couplings between a scalar field and the torsion tensor of a teleparallel connection and certain scalar field potentials, and thus proof the existence of scalarized black holes in these theories.
On the other hand, we show that it is possible to establish no-scalar-hair theorems similar to what is known in general relativity for other choices of these functions.
\end{abstract}

\maketitle

\section{Introduction}\label{sec:intro}
Black holes (BHs) play a crucial role to describe astrophysical objects. According to General Relativity (GR) they are only characterised by three parameters: their mass $M$, angular momentum $J$ and charge $Q$~\cite{Misner:1973prb,Chrusciel:2012jk,Herdeiro:2015waa}. This fact is established by the so-called no-hair theorem~\cite{Israel:1967za,Carter:1971zc} which gave rise to the conjecture that in the presence of any kind of matter, the result of a gravitational collapse would give rise to a Kerr-Newmann black hole which is fully described by these three physical quantities~\cite{Ruffini:1971bza}. For realistic astrophysical systems, the charge is thought to be very small so it is usually neglected when one is studying black hole configurations. Furthermore, by assuming that the black hole is a non-rotating one, GR predicts that the unique vacuum spherically symmetric solution is the Schwarzschild solution which characterises a black hole only by its mass~\cite{Schwarzschild:1916uq,Israel:1967wq}. 

Different studies have been trying to find black holes solutions violating the no-hair theorem. This is of course not possible in vacuum for GR, but it is possible to evade the assumptions on which the theorem is based by adding a scalar field minimally or non-minimally coupled to the gravitational sector. Several studies have found that, depending on how the coupling is implemented, the no-hair theorem is still valid or not. In~\cite{chase1970event}, it was shown that the simplest massless minimally coupled scalar field with a Lagrangian of the form $\mathcal{L}=\frac{1}{4}\lc{R}-\frac{1}{2}(\partial_\mu \psi)(\partial^\mu \psi)$ still respects the no-hair theorem. Further, adding a mass term $-\frac{1}{2} \mu^2 \psi^2$ still does not change this fact~\cite{Bekenstein:1972ny}. However, by replacing the mass term by a generic potential $\mathcal{V}(\psi)$, i.e.\ for a general Einstein-Klein-Gordon Lagrangian, one may circumvent the no-hair theorem depending on the choice of $\mathcal{V}(\psi)$~\cite{heusler1992scaling,Heusler:1996ft,Heusler:1992ss,Bekenstein:1995un,Sudarsky:1995zg,Herdeiro:2015waa}. Another way to construct scalarized black hole solutions can be found by allowing the scalar field to be complex~\cite{Pena:1997cy,Herdeiro:2015gia,Herdeiro:2014goa} or considering non-minimal couplings between the scalar field and the Ricci scalar, for example in Lagrangians like $\mathcal{L}=\frac{1}{4}F(\psi)\lc{R}-\frac{1}{2}(\partial_\mu \psi)(\partial^\mu \psi)-\mathcal{V}(\psi)$ with some specific potentials and coupling functions~\cite{Bekenstein:1975ts,Graham:2014ina,Saa:1996aw,Saa:1996qq,Sotiriou:2011dz}. Other attempts to construct scalarized black hole solutions consider a scalar field coupled to modifications and extensions of GR, motivated by the search for dark matter, dark energy and quantum gravity (see~\cite{Clifton:2011jh,Capozziello:2011et,Brihaye:2018grv,CANTATA:2021ktz,Nojiri:2017ncd, Addazi:2021xuf} for reviews). These new theories of gravity contain additional terms to the Ricci scalar (or its reformulations in terms of torsion or non-metricity) in the gravitational action. One of the most popular theories exhibiting scalarized black holes is the one where a Gauss-Bonnet invariant $\lc{G}$ is coupled to the scalar field, namely where the Lagrangian is of the form $\mathcal{L}=\frac{1}{4}\lc{R}-\frac{1}{2}(\partial_\mu \psi)(\partial^\mu \psi)+F(\psi)\lc{G}$. In these theories, it was shown that scalarized black holes exists for several choices of the non-minimal coupling function $F(\psi)$~\cite{Sotiriou:2013qea,Sotiriou:2014pfa,Brihaye:2015qtu,Silva:2017uqg,Brandelet:2017nbc,Doneva:2017bvd,Antoniou:2017acq}. The first example was provided in \cite{Sotiriou:2013qea,Sotiriou:2014pfa} assuming a linear coupling ($F(\psi) \propto \psi$). The Gauss-Bonnet invariant being a total divergence in 4$D$, such a coupling ensure that the model present a symmetry under $\psi \to \psi + c$ for a given constant $c$, dubbed as the shift-symmetry. Later on, under the assumption of a quadratic coupling ($F(\psi) \propto \psi^2$) hairy black holes were constructed in \cite{Silva:2017uqg} as the result of the so-called spontaneous scalarization process. One should also see \cite{Brihaye:2018grv} for the construction of hairy black holes extrapolating between the shift-symmetric and spontaneously scalarized types of solutions ($F(\psi) = \gamma_1 \psi + \gamma_2 \psi^2$). Some of these black holes are constructed in such a way that one has an asymptotically flat spacetime described by the Schwarzschild metric when the scalar field is vanishing. Finally, hairy black holes have also be found in theories including a non-minimal coupling between the first derivative of the scalar field and the Einstein tensor, namely $\mathcal{L} =  \lc{R} - (\eta\ g^{\mu\nu} - \beta\ \lc{G}^{\mu\nu})\partial_\mu\psi\partial_\nu\psi$, see \cite{Babichev:2013cya}. In this case, the scalar field is allowed to be time dependent even tough the spacetime is assumed to be spherically symmetric. It should be noted that all these theories can evade the no-hair theorem even without potential.

To the best of our knowledge, all the scalarised black holes found so far are described by extensions of GR based on Riemannian geometry, i.e.\ on the Levi-Civita connection with vanishing torsion and non-metricity. In this paper, we will study theories constructed in the framework of teleparallel gravity where the geometry is described by a connection with vanishing curvature and non-metricity, but non-vanishing torsion~\cite{Aldrovandi:2013wha,Bahamonde:2021gfp,Krssak:2018ywd}. It is well known that in this framework, it is possible to formulate a theory which is dynamically equivalent to GR, that is usually called the teleparallel equivalent of general relativity (TEGR)~\cite{Aldrovandi:2013wha}. The TEGR action, defined by the so called torsion scalar $T$, and the Einstein-Hilbert action differ only by a boundary term $B$ as the Ricci scalar for the Levi-Civita connection $\lc{R}$ can be expressed as $\lc{R}=-T+B$. Since $B$ is a boundary term in the action it does not influence the field equations of the theory as long as it appears linearly in the action. However, as in the standard case of Gauss-Bonnet scalar-tensor theory, the boundary term would contribute to the dynamics when coupled with the scalar field in the action.

In this work, we extend the TEGR action by considering a scalar field which is non-minimally coupled to $B$ and $T$. These theories are the so-called scalar-torsion theories (or teleparallel scalar-tensor theories), considered in the context of cosmology only with the torsion scalar already in \cite{Geng:2011aj}, later with the boundary term in~\cite{Bahamonde:2015hza}, and constructed in all generality in the series of articles~\cite{Hohmann:2018ijr,Hohmann:2018vle,Hohmann:2018dqh} as well as in the review~\cite{Bahamonde:2021gfp} (see Sec.~5.8). Exact wormholes solutions, induced by the existence of a non-trivial scalar field, were already found in~\cite{Kofinas:2015hla,Bahamonde:2016jqq}. Furthermore, it has been obtained that the PPN parameters of this theory only differs from the GR PPN parameters in $\alpha$ and $\beta$ when there is a coupling between the boundary term and the scalar field~\cite{Flathmann:2019khc}. Our main aim is to find teleparallel scalarised black holes and to investigate the existence of no-hair theorems within these theories.

This paper is organised as follows: In Sec.~\ref{sec:TPGrav} we give a brief overview of teleparallel theories of gravity and we present the scalar-torsion theory considered with its corresponding field equations. Sec.~\ref{sec:ScalarSpherical} gives the most important results of the paper: we demonstrate the existence of new scalarised black hole solutions. The section starts by presenting the field equations in spherical symmetry in Sec.~\ref{ssec:Feq} for the two possible tetrads which solve the antisymmetric field equations of the theory. Then, we analyse two main theories, namely, one which assumes only a coupling between the boundary term with the scalar field (Sec.~\ref{ssec:scalarbdryy}) and the other which only considers a coupling between the torsion scalar and the scalar field (Sec.~\ref{ssec:torsionscalarcoupling}). In these two theories we analyse the possible two tetrads and we provide new exact spherically symmetric solutions. In Sec.~\ref{sec:nohair} we discuss no-hair theorems for certain classes of theories. Finally, we provide the main conclusions of our findings in Sec.~\ref{sec:conclusion}. Throughout this paper we assume the units where $c=1$ and the metric signature is $(+---)$. Objects labeled with a $\mathring{{}}$ are constructed with help of the Levi-Civita connection of the metric defined by the tetrad. Spacetime indices are raised in lowered with the spacetime metric, Latin indices refer to the tangent spacetime indices and Greek ones to the spacetime ones.

\section{Teleparallel Gravity with non-minimally coupled scalar fields}\label{sec:TPGrav}

We briefly recall the setup of covariant teleparallel gravity as well as the action of the scalar-torsion gravity theory and its field equations, which we will solve in the next section to demonstrate the existence of scalarized black holes in scalar-torsion theories.

Standard references for teleparallel gravity are \cite{Aldrovandi:2013wha,Hohmann:2017duq,Krssak:2015oua,Bahamonde:2021gfp}. Scalar-tensor theories in teleparallel gravity have been discussed very detailed in the series of articles \cite{Hohmann:2018vle,Hohmann:2018dqh,Hohmann:2018ijr}.

\subsection{Covariant teleparallel gravity}
The fundamental variables in teleparallel gravity are a tetrad coframe $\theta^a = \theta^a{}_\mu dx^\mu$, resp. its dual $e_a = e_a{}^\mu \partial_\mu$ satisfying
\begin{equation}\label{eqn:metric}
\theta^a{}_\mu e_a{}^{\nu} = \delta^\nu_\mu,\quad \theta^a{}_\mu e_b{}^{\mu} = \delta^a_b,\quad g_{\mu\nu} = \eta_{ab}\theta^a{}_{\mu}\theta^b{}_{\nu}\,,
\end{equation}
where \(\eta_{ab} = \mathrm{diag}(1,-1,-1,-1)\) is the Minkowski metric, and a flat, metric compatible Lorentz spin connection with coefficients $\omega^a{}_{b\mu}$ which is generated by Lorentz transformation matrices $\Lambda^a{}_b$
\begin{align}
    \omega^a{}_{b\mu} = \Lambda^a{}_c\partial_\mu (\Lambda^{-1})^c{}_b\,.
\end{align}
The torsion tensor of the spin connection is given by
\begin{equation}
T^{\rho}{}_{\mu\nu} = e_a{}^{\rho}\left(\partial_{\mu}e^a{}_{\nu} - \partial_{\nu}e^a{}_{\mu} + \omega^a{}_{b\mu}e^b{}_{\nu} - \omega^a{}_{b\nu}e^b{}_{\mu} \right)\,.
\end{equation}
The spin connection is constructed such that under the action of local Lorentz transformations on the tetrads and the spin connection, the torsion tensor with only spacetime indices, as we just displayed, is invariant.

In the formulation of teleparallel geometry, i.e.\ on a manifold whose geometry is defined by tuple $(\theta^a, \omega^a{}_{b\mu})$ consisting of a tetrad and a flat, metric compatible spin connection, it is well known that one can work in the so called Weitzenb\"ock gauge. This means with a tetrad (the Weitzenb\"ock tetrad) and vanishing spin connection, i.e.\ with the tuple $(\theta^a, 0)$. Throughout this article we will work with Weitzenb\"ock tetrads.

For the scalar-torsion theories of gravity we study in Sec.~\ref{ssec:stor} we will need further ingredients. First of all, one can show that the Ricci scalar $\lc{R}$ computed with the Levi-Civita connection can be related to the so-called torsion scalar $T$ and the boundary term $B$ as follows
\begin{align}
\lc{R}= -T+B\,,\quad T=\frac{1}{2}T^{\rho\mu\nu}S_{\rho\mu\nu}\,,\quad B= 2\mathring{\nabla}_{\nu}T_{\mu}{}^{\mu\nu}\,,\label{RTB}
\end{align}
where we have defined the superpotential as
\begin{equation}\label{eq:defS}
    S_{\rho\mu\nu} = \frac{1}{2}\left(T_{\nu\mu\rho} + T_{\rho\mu\nu} - T_{\mu\nu\rho}\right) - g_{\rho\mu}T^{\sigma}{}_{\sigma\nu} + g_{\rho\nu}T^{\sigma}{}_{\sigma\mu}\,.
    \end{equation}
Eq.~\eqref{RTB} says that the Ricci scalar differs by a boundary term with respect to the torsion scalar. Thus, a Lagrangian constructed by $T$ would provide the same equations of motions that the Einstein-Hilbert action constructed from $\lc{R}$. That particular theory is called the ``Teleparallel equivalent of GR" (TEGR) since it provides the same equations as the Einstein's field equations, expressed in terms of the teleparallel quantities.

\subsection{Scalar-torsion theories of gravity}\label{ssec:stor}

To demonstrate that the teleparallel boundary term induces scalarized black holes, similarly to what the Gauss-Bonnet boundary term does, we consider scalar-torsion theories of gravity as they have been introduced in \cite{Hohmann:2018ijr}. Let $\theta = \det\theta^a{}_{\mu}$ and let $\mathcal{A}, \mathcal{B}, \mathcal{C}, \mathcal{V}$ be functions of the scalar field $\psi$. Then,
\begin{equation}\label{eq:storact}
S_{\rm g}\left[\theta^a, \omega^a{}_b, \psi\right] = \frac{1}{2\kappa^2}\int_M\left[-\mathcal{A}(\psi)T + 2\mathcal{B}(\psi)X + 2\mathcal{C}(\psi)Y - 2\kappa^2\mathcal{V}(\psi)\right]\theta\, \dd^4x\,,
\end{equation}

defines a scalar-torsion theory of gravity with $X$ being the kinetic term of the scalar field and $Y$ a derivative coupling term defined by \begin{align}\label{eq:defX}
    X &= -\frac{1}{2}g^{\mu\nu}\partial_\mu \psi \partial_\nu \psi\,, \quad
    Y = g^{\mu\nu}T^{\rho}{}_{\rho\mu}\partial_\nu\psi\,.
    \end{align}
 It has been demonstrated that this action is equivalent to the action
\begin{equation}\label{C}
S_{\rm g}\left[\theta^a,\omega^a{}_b, \psi\right] = \frac{1}{2\kappa^2}\int_M\left[-\mathcal{A}(\psi)T + 2\mathcal{B}(\psi)X - \tilde{\mathcal{C}}(\psi)B - 2\kappa^2\mathcal{V}(\psi)\right]\theta\, \dd^4x\,,
\end{equation}
up to a boundary term, when one chooses the coupling function $\mathcal{C}(\psi) = \partial_\psi \tilde{\mathcal{C}}(\psi)$. In the following we always assume this equivalence. 

This gravitational part of the action is coupled to matter actions $S_{\rm m}$, which we assume to depend only on the tetrad via the metric, not on the spin connection or the torsion. Variation of the total action $S = S_{\rm g} + S_{\rm m}$ with respect to the tetrad, then leads to gravitational field equations of the form
\begin{align}
    E_a{}^\mu = \kappa^2 \Theta_a{}^\mu\,.
\end{align}
Multiplying with the components of tetrad and lowering the indices with the metric yields equations of the type $E_{\mu\nu} = \kappa^2 \Theta_{\mu\nu}$, which decay into symmetric $E_{(\mu\nu)} = \kappa^2 \Theta_{(\mu\nu)}$ and antisymmetric $E_{[\mu\nu]} = 0$ part. It can be shown that the atisymmetric equations of the variation of the action with respect to the tetrad are identical to the equations one obtains when one varies the action with respect to the spin connection \cite{Hohmann:2017duq,Bahamonde:2021gfp}. Since we are working with the Weitzenb\"ock tetrad, the tetrad alone must satisfy both, the symmetric and the antisymmetric field equations. 

The general scalar-torsion theories of gravity \eqref{eq:storact} and \eqref{C} are dynamically equivalent to general relativity with a minimally coupled scalar field by choosing either $(\mathcal{A}= \tilde{\mathcal{C}} = \alpha, \mathcal{B}=\beta)$ or $(\mathcal{A} = \alpha, \mathcal{B}=\beta, \tilde{\mathcal{C}} = 0)$, for constants $\alpha$ and $\beta$. The first case yields the standard Einstein-Klein-Gordon action, while the second differs from that by a boundary term, and is the standard TEGR action supplemented by a minimally coupled scalar field.

The field equations of the general theory \eqref{eq:storact} have been derived in \cite{Hohmann:2018ijr}. Hereafter we will assume that $\mathcal{B}(\psi)=\beta$. By taking variations with respect to the tetrads, the symmetric part is given by
\begin{multline}\label{eqn:clafeqtets}
\left(\mathcal{A}'(\psi) + \mathcal{\tilde{C}}'(\psi)\right)S_{(\mu\nu)}{}^{\rho}\psi_{,\rho} + \mathcal{A}(\psi)\left(\mathring{R}_{\mu\nu} - \frac{1}{2}\mathring{R}g_{\mu\nu}\right) + \left(\frac{1}{2}\beta- \mathcal{\tilde{C}}''(\psi)\right)\psi_{,\rho}\psi_{,\sigma}g^{\rho\sigma}g_{\mu\nu}\\
- (\beta- \mathcal{\tilde{C}}''(\psi))\psi_{,\mu}\psi_{,\nu} + \mathcal{\tilde{C}'}(\psi)\left(\mathring{\nabla}_{\mu}\mathring{\nabla}_{\nu}\psi - \mathring{\square}\psi g_{\mu\nu}\right) + \kappa^2\mathcal{V}(\psi)g_{\mu\nu} = \kappa^2\Theta_{\mu\nu}\,,
\end{multline}
where here and in the following a $'$ denote the derivative of a function with respect to its argument, while the antisymmetric part of the field equation becomes
\begin{align}\label{eq:asym}
   \left(\mathcal{A}'(\psi)+ \tilde{\mathcal{C}}'(\psi)\right) T^{\rho}{}_{[\mu\nu}\psi_{,\rho]} = 0\,.
\end{align}
Variations with respect to the scalar field provides us the modified Klein-Gordon equation:
\begin{equation}\label{eqn:clafeqscal}
\frac{1}{2}\mathcal{A}'(\psi)T+ \frac{1}{2}\mathcal{\tilde{C}}'(\psi)B  - \beta\mathring{\square}\psi
+ \kappa^2\mathcal{V}'(\psi) = \kappa^2\epsilon\Theta\,,
\end{equation}
where $\epsilon$ provides a non-minimal coupling between the scalar field and the matter content. 

To demonstrate that scalarized black holes exist in the scalar-torsion theories of gravity we just displayed, we will focus on the equations in vacuum, i.e.\ for $\Theta_{\mu\nu} = 0$. Observe that the symmetric field equation \eqref{eqn:clafeqtets} can also be understood as Einstein equations with effective energy momentum tensor $\Theta^{(\psi)}{}_{\mu\nu}$ sourced by the superpotential of the torsion and the scalar field, by writing \eqref{eqn:clafeqtets} as
\begin{align}\label{eq:symGR}
\begin{split}
\mathring{R}_{\mu\nu} - \frac{1}{2}\mathring{R}g_{\mu\nu}&=-\frac{1}{\mathcal{A}(\psi)}\Big[\left(\mathcal{A}'(\psi) + \mathcal{\tilde{C}}'(\psi)\right)S_{(\mu\nu)}{}^{\rho}\psi_{,\rho}  + \left(\frac{1}{2}\beta- \mathcal{\tilde{C}}''(\psi)\right)\psi_{,\rho}\psi_{,\sigma}g^{\rho\sigma}g_{\mu\nu}\\
&- (\beta- \mathcal{\tilde{C}}''(\psi))\psi_{,\mu}\psi_{,\nu} + \mathcal{\tilde{C}'}(\psi)\left(\mathring{\nabla}_{\mu}\mathring{\nabla}_{\nu}\psi - \mathring{\square}\psi g_{\mu\nu}\right) + \kappa^2\mathcal{V}(\psi)g_{\mu\nu}\Big]=: \Theta^{(\psi)}{}_{\mu\nu}\,.
 \end{split}
\end{align}
This interpretation of the field equations helps us to identify an effective energy $\mathcal{E}_\psi$ of the scalar field to characterise its behaviour. Assuming the existence of a timelike killing vector field $\xi=\xi^\mu\partial_\mu$, $\mathcal{E}_\psi$ is given by \cite{CarollGR,WaldGR}
\begin{equation}\label{eq:Epsi}
    \mathcal{E}_\psi \coloneqq \int_{\mathcal{H}} \Theta^{(\psi)}{}^{\mu}{}_{\nu}\  \xi^\nu n_\mu \sqrt{-\det(h)} \dd^3x\,,
\end{equation}
where $\mathcal{H}$ is a spacelike hypersurface with induced negative definite metric $h$ and timelike conormal field $n = n_\mu \dd x^\mu$. 

From Sec.~\ref{sec:ScalarSpherical} on we will consider spacetime geometries with a metric of the form
\begin{align}\label{eq:met}
    \dd s^2 = A^2 \dd t^2 - \frac{C^2}{A^2}\dd r^2-r^2 \dd \Omega^2\,,
\end{align}
where $\dd \Omega^2=r^2(\dd\theta^2+\sin^2\theta \dd\varphi^2)$. The slightly unusual parametrization in terms of the functions $A=A(r)$ and $C=C(r)$ here is chosen to ensure non-degeneracy of the metric at the black hole horizon. For the analysis of the solutions we present in the following, we remark that such metrics are asymptotically flat if and only if $A(r)\overset{r\to\infty}{\longrightarrow} c<\infty$ and $C(r)\overset{r\to\infty}{\longrightarrow}c$, where $c$ is a constant, so that $C A^{-1}\overset{r\to\infty}{\longrightarrow} 1$.

The metric clearly possesses a timelike Killing vector field $\xi = \partial_t$ and we can express the energy of the scalar field in the black hole exterior region, i.e.\ on a spacelike hypersurface $\mathcal{H}_{\text{ext}}$ that is defined by $x^0=t=\textrm{const.}$ and $x^1=r \ge r_h$, $r_h$ being the radius of the event horizon, with conormal $n=A \dd t$ and induced metric $h = - C^2 A^{-2}\dd r^2-r^2\dd\Omega^2$, as
\begin{align}
    \mathcal{E}_\psi = \int_{r_h}^\infty \int_{S^2} \Theta^{(\psi)}{}^{t}{}_{t} C r^2 \dd r\, \dd\Omega = 4\pi  \int_{r_h}^\infty \rho_\psi(r) C r^2 \dd r\,,\label{energy}
\end{align}
where $\rho_\psi (r) \coloneqq \Theta^{(\psi)}{}^{t}{}_{t}(r)$ will be interpreted as the (effective) scalar field energy density.

\section{Scalarized teleparallel spherically symmetric static black holes}\label{sec:ScalarSpherical}
The field equations in spherical symmetry can be derived for the general Lagrangian \eqref{C}. Based on what is know from teleparallel theories of gravity in spherical symmetry \cite{Pfeifer:2022txm,Bahamonde:2021srr,Hohmann:2019nat}, it is straightforward to find two classes of tetrads which solve the antisymmetric fields equations, a real and a complex one. For these tetrads, one can display the remaining symmetric field equations, which is what we do first in Sec.~\ref{ssec:Feq}. 

Solving them in general, for arbitrary coupling functions $\mathcal{A}(\psi)$ and $\tilde{\mathcal{C}}(\psi)$ is not possible. However, for specific choices of theories, with certain fixed values for these coupling functions we find scalarized black hole solutions. In Sec.~\ref{ssec:scalarbdryy} we consider only a coupling between the scalar field and the boundary term ($\mathcal{A}(\psi)=\alpha$), while in Sec.~\ref{ssec:torsionscalarcoupling} only a coupling between the torsion scalar and the scalar field ($\tilde{\mathcal{C}}(\psi)=0$) is considered. 

In both classes of scalar-torsion extensions of general relativity, scalarized black holes exist for suitable choices of the coupling functions.

In the next section \ref{sec:nohair} we will discuss the existence and non-existence of no-hair theorems for scalar-torsion theories.

\subsection{The field equations for the real and complex tetrad}\label{ssec:Feq}
In~\cite{Bahamonde:2021srr} it was found that for a generic $f(T,B,\psi,X)$ gravity theory, the antisymmetric field equations~\eqref{eq:asym} are solved for only two possible classes of tetrads (in the Weitzenb\"ock gauge). This statement assumes that both the metric and teleparallel connection respect spherical symmetry. The first tetrad is real and it is described by~ \cite{Tamanini:2012hg,Hohmann:2019nat}
\begin{align}
    \theta^{(1)}{}^a{}_\mu =\left(
    \begin{array}{cccc}
    A & 0 & 0 & 0 \\
    0 & C A^{-1} \cos\phi \sin\theta & r \cos\phi \cos\theta  & - r \sin\phi \sin\theta  \\
    0 & C A^{-1} \sin\phi \sin\theta  & r \sin\phi \cos\theta  &  r \cos\phi \sin\theta \\
    0 & C A^{-1} \cos\theta & - r \sin\theta & 0 \\
    \end{array}
    \right)\label{tetrad}\,,
\end{align}
whereas the second one is complex and it is given by~\cite{Bahamonde:2021srr}
\begin{align}
    \theta^{(2)}{}^a{}_\mu =\left(
\begin{array}{cccc}
 0 & \frac{i C}{A} & 0 & 0 \\
 i A \sin \theta \cos\phi & 0 & -r \sin\phi & -r \sin \theta \cos\theta \cos\phi \\
 i A \sin \theta \sin\phi & 0 & r \cos\phi & -r \sin \theta \cos\theta \sin\phi \\
 i A \cos\theta & 0 & 0 & r \sin ^2\theta  \\
\end{array}
\right)\label{tetrad2}\,.
\end{align}
The remaining equations, which need to be integrated, are the symmetric ones \eqref{eqn:clafeqtets} and the scalar field equation \eqref{eqn:clafeqscal}. There will be two different sets of equations depending on the choice of the tetrad which satisfy the antisymmetric field equations. In addition, we require the scalar field to only depend on the radial coordinate to also respect spherical symmetry $\psi=\psi(r)$. 

For the real tetrad~\eqref{tetrad} the torsion scalar, vector part of torsion and boundary term become
\begin{eqnarray}
    T&=&\frac{2 (A-C) \left(2 r A'+A-C\right)}{r^2 C^2}\,,\label{T1}\\
     T_\mu&=&\Big(0,-\frac{ A'}{ A}-\frac{2 }{r }+\frac{2 C}{r A},0,0\Big)\,,\label{Tmu1}\\
      B&=&\frac{2 \left(r C A' \left(r A'-2 C\right)+A \left(-r^2 A' C'+r C \left(r A''+6 A'\right)-2 C^2\right)+2 A^2 \left(C-r C'\right)\right)}{r^2 C^3}\,,\label{B1}
\end{eqnarray}
while for the complex tetrad~\eqref{tetrad2} these quantities are
\begin{eqnarray}
    T&=&\frac{2 \left(2 r A A'+A^2+C^2\right)}{r^2 C^2}\,,\label{T2}\\
     T_\mu&=&\Big(0,-\frac{A'}{A}-\frac{2}{r},0,0\Big)\,,\label{Tmu2}\\
      B&=&\frac{2 \left(r^2 C A'^2+r A \left(C \left(r A''+6 A'\right)-r A'C'\right)+2 A^2 \left(C-r C'\right)\right)}{r^2 C^3}\,.\label{B2}
\end{eqnarray}
Notice that as expected, both tetrads give the same form of the Ricci scalar $\lc{R}$
\begin{eqnarray}\label{eq:ricci}
    \lc{R}=-T+B = -\frac{2 A \left(r A'+2 A\right) C'}{r C^3}+\frac{2 \left(r^2 A'^2+r A \left(r A''+4 A'\right)+A^2\right)}{r^2 C^2}-\frac{2}{r^2}\,.
\end{eqnarray}
One can immediately notice some differences between the real and complex tetrad. First of all, the torsion scalar and boundary term only vanish for the real tetrad when one considers a  Minkowski metric. Further, the vectorial part of torsion $T_\mu$ in the complex tetrad does not depend on $C$ whereas in the real tetrad it does. Another interesting point to remark is that in the Schwarzschild case, the torsion scalar and boundary term become $4/r^2$ for the complex tetrad and therefore they are regular at the horizon $r=2M$, while, for the real tetrad, these scalars diverge at $r=2M$. 

\subsubsection{The field equations for the real tetrad}\label{sec:real}
Employing the first tetrad \eqref{tetrad} in the symmetric vacuum field equations, \eqref{eqn:clafeqtets} and \eqref{eqn:clafeqscal} with $\Theta_{\mu\nu}=0$, leads to
\begin{subequations}\label{eq:eomrealtet}
\begin{eqnarray}
E^{(1)}_{tt}=0&=&\mathcal{A}(\psi)\frac{A^2 \left(-2 r A C A'-A^2 \left(C-2 r C'\right)+C^3\right) }{r^2 C^3}+\mathcal{\tilde{C}}'(\psi) \left(\frac{A^4 \psi''}{C^2}-\frac{A^3 \psi' \left(-r C A'+r A C'-2 C^2\right)}{r C^3}\right)\nonumber\\
&&-\frac{2 A^3 (A-C) \psi' \mathcal{A}'(\psi)}{r C^2}+\frac{A^4 \psi'^2  \mathcal{\tilde{C}}''(\psi) }{C^2}-\frac{A^4 \psi'^2 \beta}{2 C^2}+\kappa ^2 A^2 \mathcal{V}(\psi)\,,\label{eq:ttF}\\
E^{(1)}_{rr}=0&=&-\frac{\left(2 r A A'+A^2-C^2\right) \mathcal{A}(\psi)}{r^2 A^2}+\left(\frac{A'}{A}+\frac{2}{r}\right) \psi' \mathcal{\tilde{C}}'(\psi) +\frac{\kappa ^2 C^2 \mathcal{V}(\psi)}{A^2}+\frac{1}{2} \beta  \psi'^2\,,\label{eq:rrF}\\
E^{(1)}_{\theta\theta}=0&=&-\frac{A \psi' \left(r A'+A-C\right) \mathcal{A}'(\psi) }{r C^2}+\mathcal{\tilde{C}}'(\psi)  \left(\frac{A \psi' \left(r C A'-r A C'+C^2\right)}{r C^3}+\frac{A^2 \psi''}{C^2}\right)\nonumber\\
&&+\frac{\mathcal{A}(\psi)\left(-r C A'^2+A \left(r A' C'-C \left(r A''+2 A'\right)\right)+A^2 C'\right)}{r C^3}+\frac{A^2 \psi'^2  \mathcal{\tilde{C}}''(\psi) }{ C^2}-\frac{A^2 \psi'^2 \beta }{2 C^2}+\kappa ^2 \mathcal{V}(\psi)\,,\label{eq:ththF}\\
E^{(1)}_{\psi}=0&=&\frac{(A-C) \left(2 r A'+A-C\right) \mathcal{A}'(\psi) }{r^2 C^2}+\frac{\beta  A \psi' \left(2 r C A'+A \left(2 C-r C'\right)\right)}{r C^3}+\frac{\beta  A^2 \psi''}{C^2}+\kappa ^2 \mathcal{V}'(\psi)\nonumber\\
&&+\frac{\mathcal{\tilde{C}}'(\psi)  \left(r C A' \left(r A'-2 C\right)+A \left(-A'r^2  C'+r C \left(r A''+6 A'\right)-2 C^2\right)+2 A^2 \left(C-r C'\right)\right)}{r^2 C^3}\,.\label{eq:psiF}
\end{eqnarray}
\end{subequations}

\subsubsection{The field equations for the complex tetrad}\label{sec:complex}
For the complex tetrad~\eqref{tetrad2}, the Eqs.~\eqref{eqn:clafeqtets} and \eqref{eqn:clafeqscal} in vacuum become
\begin{subequations}\label{eq:eomcmplxtet}
\begin{eqnarray}
E^{(2)}_{tt}=0&=&\mathcal{A}(\psi)\frac{A^2  \left(-2 r A C A'-A^2 \left(C-2 r C'\right)+C^3\right)}{r^2 C^3}+\mathcal{\tilde{C}}'(\psi) \frac{A^3 \left(C \left(A' \psi'+A \psi''\right)-A C' \psi'\right)}{C^3}-\frac{2 A^4 \psi' \mathcal{A}'(\psi)}{r C^2}\nonumber\\
&&+\frac{A^4 \psi'^2 \mathcal{\tilde{C}}''(\psi)}{C^2}-\frac{\beta  A^4 \psi'^2}{2 C^2}+\kappa ^2 A^2 \mathcal{V}(\psi)\,,\label{eq:ttF2}\\
E^{(2)}_{rr}=0&=&\frac{\mathcal{A}(\psi) \left(-2 r A A'-A^2+C^2\right)}{r^2 A^2}+\left(\frac{A'}{A}+\frac{2}{r}\right) \psi' \mathcal{\tilde{C}}'(\psi)+\frac{\kappa ^2 C^2 \mathcal{V}(\psi)}{A^2}+\frac{1}{2} \beta  \psi'^2\label{eq:rrF2}\,,\\
E^{(2)}_{\theta\theta}=0&=&-\mathcal{A}'(\psi)\frac{A \left(r A'+A\right) \psi' }{r C^2}+\mathcal{\tilde{C}}'(\psi)\frac{ A \left(C \left(A' \psi'+A \psi''\right)-A C' \psi'\right)}{C^3}\nonumber\\
&&+\frac{\mathcal{A}(\psi) \left(-r C A'^2+A \left(r A' C'-C \left(r A''+2 A'\right)\right)+A^2 C'\right)}{r C^3}+\frac{A^2 \psi'^2 \mathcal{\tilde{C}}''(\psi)}{C^2}-\frac{\beta  A^2 \psi'^2}{2 C^2}+\kappa ^2 \mathcal{V}(\psi)\label{eq:ththF2}\,,\\
E^{(2)}_{\psi}=0&=&
\frac{\left(2 r A A'+A^2+C^2\right) \mathcal{A}'(\psi)}{r^2 C^2}+\frac{\beta  A \left(2 r C A' \psi'+A \left(C \left(r \psi''+2 \psi'\right)-r C' \psi'\right)\right)}{r C^3}\nonumber\\
&&+\frac{\mathcal{\tilde{C}}'(\psi) \left(r^2 C A'^2+r A \left(C \left(r A''+6 A'\right)-r A' C'\right)+2 A^2 \left(C-r C'\right)\right)}{r^2 C^3}+\kappa ^2 \mathcal{V}'(\psi)\label{eq:psiF2}\,.
\end{eqnarray}
\end{subequations}

\subsubsection{Remarks on the field equations}
A first observation is that the $rr$-component and the modified Klein-Gordon equation for the two tetrads differ only by
\begin{eqnarray}
E^{(1)}_{rr}-E^{(2)}_{rr}&=&-\frac{2 \left(r A'+A\right) }{r^2 C}\left(\mathcal{A}'(\psi)+\mathcal{\tilde{C}}'(\psi)\right)\,,\\
E^{(1)}_{\psi}-E^{(2)}_{\psi}&=&-\frac{(C-1) C }{r^2 A^2}\left(\mathcal{A}(\psi)+\kappa ^2 r^2 \mathcal{V}(\psi)\right)\,,
\end{eqnarray}
and the difference between the other components contain much more terms. This means that when $C=1$ ($g_{rr}=-1/g_{tt}$), the modified Klein-Gordon equations become the same for the two tetrads.

Moreover ,the systems~\eqref{eq:ttF}-\eqref{eq:psiF} and~\eqref{eq:ttF2}-\eqref{eq:psiF2} there are only four equations since we did not display $E_{\phi\phi}$, which is directly related to $E_{\theta\theta}$ via $E_{\phi\phi}= \sin^2\theta E_{\theta\theta}$, due to the spherical symmetry we are considering. 
We like to remark that these four equations for the three variables $A$, $C$, and $\psi$ are not independent. For example, for the first tetrad, one can notice that Eq.~\eqref{eq:ththF} is solved, by employing $A'$ (and its derivative) from \eqref{eq:rrF}, $C'$ from \eqref{eq:ttF} and $\psi''$ from \eqref{eq:psiF}.

Thus generically, we have three equations for three unknowns $A,C$ and $\psi$ as function of $r$. Hence, in principle, there should exist solutions for sufficiently regular choices of the couplings $\mathcal{A}$, $\mathcal{\tilde C}$ and the potential $\mathcal{V}$. In this paper, we will focus on explicit solutions that admit a closed form in terms of either usual or special functions. It is worth noting that surely the spectrum of solutions in general is way larger. Indeed, the systems \eqref{eq:eomrealtet} or \eqref{eq:eomcmplxtet} are highly non-linear systems of differential equations. In general it will not be possible (i.e. for any choice of $\mathcal{A}(\psi)$, $\mathcal{V}(\psi)$ and $\mathcal{\tilde{C}}(\psi)$) to find solutions in closed form. The existence of such solutions is a remarkable property achieved for some specific choices of the functions $\mathcal{A}(\psi)$, $\mathcal{V}(\psi)$ and $\mathcal{\tilde{C}}(\psi)$. The analysis in this paper could then be completed by a numerical investigation of the spectrum of solutions for some specific choices of these functions. We keep such an analysis for later work.

In the next sections we find non-trivial analytic solutions for the field equations for specific scalar-torsion extensions of general relativity.

\subsection{\texorpdfstring{$\mathcal{A}(\psi)=\alpha$\,:}\, only non-minimal coupling between the scalar field and the boundary term}\label{ssec:scalarbdryy}
It is well known that non-minimally coupled scalar to boundary terms built from the metric and the curvature of the Levi-Civita connections, such as for example most famously the Gauss-Bonnet boundary term, lead to scalar-tensor theories of gravity which have scalarized black hole solutions~\cite{Silva:2017uqg,Sotiriou:2013qea,Doneva:2017bvd,Antoniou:2017acq}.

In this section we consider a coupling between the scalar field and the teleparallel boundary term $B$, which is the difference between the torsion scalar and the Ricci scalar. We find that such theories also allow for several classes of scalarized solutions. Among them is the Schwarzschild-(anti)-de-Sitter (S-(A)dS) spacetime equipped with a non-trivial scalar field.

\subsubsection{Analytical analysis of the field equation for the real tetrad}\label{sssec:real1}

For the real tetrad we discuss several types of analytical solutions to boundary-term non-minimal coupling scalar-torsion gravity.

Let us first explore the case $\alpha=0$ which is a limiting case when the theory does not have a GR limit. When we solve Eqs.~\eqref{eq:ttF}-\eqref{eq:rrF} for the coupling functions $\tilde{\mathcal{C}}'$ and $\tilde{\mathcal{C}}''$ and then we replace those expressions into~\eqref{eq:ththF} we get $\mathcal{V}(\psi)=-\frac{\beta  A^2 \psi '^2}{2 \kappa ^2 C^2}$. By replacing this form of the potential back in~\eqref{eq:rrF} we immediately get that $\tilde{\mathcal{C}}'\psi'=0$  which is only true for the trivial case when either the scalar field is a constant or the contribution from the boundary term disappears in the equations ($\tilde{\mathcal{C}}=\textrm{const}$). Thus we found that, as expected since the torsion scalar is missing in the action, this case  gives only trivial solutions of a non-dynamical scalar field. Another limiting case is when we set $A(r)=A_0/r^2$ which eliminates the coupling function $\mathcal{\tilde{\mathcal{C}}}'(\psi)$ in~\eqref{eq:rrF}. For this case, again there are only trivial solutions to the system.

Non-trivial solutions can be found by setting without loss of generality $\alpha = 1\neq0$, and $A\neq A_0/r^2 $. From now on, we will assume those conditions. We can solve~\eqref{eq:ttF}-\eqref{eq:ththF} for $\mathcal{\tilde{C}}'(\psi), \mathcal{\tilde{C}}''(\psi)$ and the potential $\mathcal{V}(\psi)$ and then replace them in~\eqref{eq:psiF} yielding
\begin{eqnarray}
0&=&-  r^3 A^2 A''' C^2+  r^2 A^2 C \left(r A'-A\right) C''+3   r^2 A^2 \left(A-r A'\right) C'^2+  r^2 A C C' \left(2 A' \left(r A'+C\right)+A \left(3 r A''-2 A'\right)\right)\nonumber\\
&&+C^2 \Big[  \left(-2 r^2 C A'^2+r C^2 A'+r^3 A'^3-2 C^3\right)+A^2 \left(  r \left(r A''+A'\right)+C \left(2  -\beta  r^2 \psi '^2\right)\right)\nonumber\\
&&+  A \left(-2 r^2 C A''+r^2 A' \left(A'-2 r A''\right)+3 C^2\right)-3   A^3\Big]\,. \label{eqGen1}
\end{eqnarray}
This equation is thus a general necessary relationship that must be always true for any form of the potential and any form of the coupling function. The above equation admits several interesting solutions, but is involved to be solved in all generality. We discuss two classes of solutions.
\begin{itemize}[leftmargin=*]
    \item A first class of solutions can be found in the case $C(r)=1$ and $\mathcal{V}(\psi)=0$. Solving~\eqref{eq:rrF}-\eqref{eq:ththF} for $\mathcal{\tilde{C}}',\mathcal{\tilde{C}}''$ and replacing them in~\eqref{eq:ttF}, gives a remaining equation
    \begin{eqnarray}
        0&=&2   \left(r^3 A'^3+r A'-1\right)+A^2 \left(2  +4   r^2 A''-2   r A'-\beta  r^2 \psi '^2\right)-4   A^3\nonumber\\
        &&+2   A \left(2 r^2 A'^2+r A' \left(r^2 A''+2\right)+2\right)\,.
    \end{eqnarray}
    An exact solution of this equation is
    \begin{align}
        \dd s^2&=\Big(1-K r\Big)^{2}\dd t^2-\Big(1-K r\Big)^{-2}\dd r^2-r^2\dd \Omega^2\,,\label{solmetric1}\\
       \psi(r)&=\sqrt{-\frac{6 }{\beta }} \log (1-K r)\,,\quad \mathcal{\tilde{C}}(\psi)=\sqrt{\frac{-2\beta}{3}} \psi\,,\quad \mathcal{V}(\psi)=0\,.\label{solpsi}
    \end{align}
    To obtain this solution we have first found the scalar field and then we have inverted it from $\psi=\psi(r)$ to $r=r(\psi)$. After doing this, we can write down the remaining field equations depending on $\psi$ (not $r$). Then, the coupling function and potential can be easily found by solving the remaining field equations. This procedure will be also used in the next sections to find the form of the potential and the coupling function. This solution is non-asymptotically flat and contains a horizon at $r_h=1/K$. Since we assumed that the scalar field and the coupling function are real, the kinetic parameter must be $\beta<0$. 
    
    \item A second class of solutions we like to display has $C(r)=-\frac{1}{2\sqrt{1-\Lambda r^2}}$ and the metric behaves as 
    \begin{eqnarray}
        \dd s^2&=&\Big(1-\Lambda r^2\Big)\dd t^2-\frac{1}{4}\Big(1-\Lambda r^2\Big)^{-2}\dd r^2-r^2\dd \Omega^2\,,\label{solmetric2B}\\
        \psi(r)&=&\sqrt{\frac{6}{\beta}}\log r\,.\label{solpsi2}
    \end{eqnarray}
    Note that $C(r)^2$ appears in the metric and $C(r)$ in the tetrad so that its sign could only affects the tetrad. Nevertheless, $C(r)=+\frac{1}{2\sqrt{1-\Lambda r^2}}$ does not have the same solution as above. The reason comes from the fact that $C(r)$ appears linearly in some of the equations and thus, its sign affects the form of the solution even though the metric is unaffected. 
    By replacing the above solution in the field equations we find that in order to obtain the potential and the coupling function one needs to invert $\psi=\psi(r)$ to $r=r(\psi)$ using~\eqref{solpsi2}. Then, it is easy to solve the remaining part of the system~\eqref{eq:ttF}-\eqref{eq:psiF} leading to the following form of the potential and coupling function
    \begin{eqnarray}
        \kappa^2\mathcal{V}(\psi)&=&3   \left(16 \Lambda ^2 e^{\sqrt{\frac{2\beta}{3}}\psi}   +e^{-\sqrt{\frac{2\beta}{3}}\psi}   -14 \Lambda \right)\,,\\
        \mathcal{\tilde{C}}(\psi)&=&-\frac{1}{4} \left( \sqrt{6\beta } \psi +5   \log \left(1-\Lambda  e^{\sqrt{\frac{2\beta}{3}}\psi }  \right)\right)\,.\label{solpsi33}
    \end{eqnarray}
    This solution represents again a non-asymptotically flat spherically symmetric solution, which behaves similarly as a S-(A)dS spacetime without a mass or at large $r$ and $g_{rr}\neq -1/g_{tt}$. Contrary to the solution~\eqref{solpsi}, the above solution requires $\beta>0$ to ensure that both the coupling function and the scalar field are real. We like to point out that for $\Lambda>0$ this solution cannot be interpreted as a black hole since the determinant of the metric, proportional to $C(r)$, diverges at the horizon. For $\Lambda<0$ there is no horizon at all.
\end{itemize}

Thus we found two classes of non-asymptotically flat scalarized solutions for a scalar field non-minimally coupled to the teleparallel boundary term in the case of a real tetrad. 

\subsubsection{Analytical analysis of the field equation for the complex tetrad}
\label{sssec:complex1}
In this section we will find exact solutions for the complex tetrad when one assumes that there is only a coupling between the boundary term and the scalar field. If we replace Eqs.~\eqref{eq:ttF2} and~\eqref{eq:rrF2} into~\eqref{eq:ththF2} we find that the metric functions must obey the following differential equation
\begin{eqnarray}
C'(r)&=&\frac{C \left(r^2 A A''+r^2 A'^2-A^2+C^2\right)}{r A \left(r A'-A\right)}\,.\label{eqC}
\end{eqnarray}
This equation is, analogously what we found for the real tetrad, a necessary condition that has to hold, independently of the theory considered. It cannot easily be solved without making further assumptions.

We investigate two different main cases. 
\begin{itemize}[leftmargin=*]
    \item First we set again $C=1$, which implies from \eqref{eqC} that the metric becomes
    \begin{equation}
    \dd s^2=\Big(1-\frac{2 M}{r}-\Lambda  r^2\Big)\dd t^2-\Big(1-\frac{2 M}{r}-\Lambda  r^2\Big)^{-1}\dd r^2-r^2\dd \Omega^2\,.\label{metricCase1}
    \end{equation}
    This means that for any form of the potential or coupling function this case yields the unique solution for the metric, which is given by a S-(A)dS metric. This result goes in a similar directions as what was found for $f(T)$-gravity in~\cite{Bahamonde:2021srr}, where it was shown that for any $f(T)$ gravity theory, the condition $g_{tt}=-1/g_{rr}$ imposes the metric to be Schwarzchild-de-Sitter.
    
    Having found the metric which solves the necessary condition~\eqref{eqC} we still need to solve the remaining field equations. There are different ways of solving them since one can assume either a form of $\mathcal{V}(\psi),\, \mathcal{\tilde{C}}(\psi)$ or even the form of the scalar field $\psi$.
    
    To demonstrate the existence of scalarized S-(A)dS spacetimes we choose the coupling function to be $\mathcal{\tilde{C}}(\psi)=(\mathcal{\tilde{C}}_0/p)\, \psi^p$, which makes 
    Eqs.~\eqref{eq:ttF2}-\eqref{eq:rrF2}
    \begin{eqnarray}
        2 \kappa ^2 r^2 \psi\, \mathcal{V}(\psi)&=&2 \mathcal{\tilde{C}}_0 \left(3 M+3 \Lambda  r^3-2 r\right) \psi^p \psi '+r \psi \left(\beta  \left(2 M+\Lambda  r^3-r\right) \psi'^2-6   \Lambda  r\right)\,,\label{Vcase1}\\
        0&=& \mathcal{\tilde{C}}_0 (p-1) r \psi^p \psi '^2+ \mathcal{\tilde{C}}_0 \psi^{p+1} \left(r \psi ''-2 \psi '\right)-\beta  r \psi^2 \psi '^2\,.\label{phicase1}
    \end{eqnarray}
    When we now fix $p$, we only need to solve equation \eqref{phicase1} for $\psi$ to obtain complete solutions of the field equations with a potential given by \eqref{Vcase1}, where the difficulty lies in expressing the potential as function of $\psi$, instead of as a function of $r$. We display solutions for several choices:
    \begin{itemize}[leftmargin=*]
        \item $p=2$, $\mathcal{\tilde{C}}_0\neq \beta/2$: 
        \begin{eqnarray}
            \psi(r)&=&\left(u\, r^3 +v\right)^{\frac{\mathcal{\tilde{C}}_0}{u}}\,,\quad \mathcal{\tilde{C}}(\psi)=\frac{1}{2}\mathcal{\tilde{C}}_0 \psi^2\,,\quad u=2\mathcal{\tilde{C}}_0-\beta\,,\quad v=3 \mathcal{\tilde{C}}_0\psi_1\,,\label{sol1b}\\
            \kappa^2 \mathcal{V}(\psi)&=&-3 \alpha  \Lambda +\frac{18 \mathcal{\tilde{C}}_0^3 \psi ^{2-\frac{u}{\mathcal{\tilde{C}}_0}} (M u-\Lambda  v)}{u^2}-\frac{9 \mathcal{\tilde{C}}_0^2 v \beta \psi ^{2-\frac{2 u}{\mathcal{\tilde{C}}_0}} (2 M u-\Lambda  v)}{2 u^2}+\frac{9 \mathcal{\tilde{C}}_0^2 \Lambda  \psi ^2 (2 \mathcal{\tilde{C}}_0+u)}{2 u^2}\nonumber\\
            &&+\frac{3 \mathcal{\tilde{C}}_0^2 \left(3 v \beta \psi ^{2-\frac{2 u}{\mathcal{\tilde{C}}_0}}-(6 \mathcal{\tilde{C}}_0+u) \psi ^{2-\frac{u}{\mathcal{\tilde{C}}_0}}\right) \sqrt[3]{\psi ^{u/\mathcal{\tilde{C}}_0}-v}}{2 u^{4/3}}\,. \label{sol1a}
        \end{eqnarray}
        \item $p=2$, $\mathcal{\tilde{C}}_0= \beta/2$:
        \begin{eqnarray}
            \psi(r)&=&e^{\frac{r^3 \psi_1}{3}}\,,\quad \mathcal{\tilde{C}}(\psi)=\frac{\beta}{4} \psi^2\,,\label{sol2b}\\
            \kappa^2 \mathcal{V}(\psi)&=&-3 \alpha  \Lambda+\frac{1}{2} \beta  \psi^2 
            \left(9 \Lambda  (\log (\psi )+1) \log (\psi) + 3 M \psi_1 (2 \log (\psi)+1)-\sqrt[3]{3} \psi_1^{2/3} (3 \log (\psi)+2) \sqrt[3]{\log (\psi)}\right)\,.\nonumber\\
            &&\label{sol2a}
        \end{eqnarray}
        \item $p=1$: 
        \begin{eqnarray}
            \psi(r)&=&-\frac{ \mathcal{\tilde{C}}_0 }{\beta }\log \left(3  \mathcal{\tilde{C}}_0 \psi_1-\beta  r^3\right)\,,\quad \mathcal{\tilde{C}}(\psi)= \mathcal{\tilde{C}}_0 \psi\,,\\
            \kappa^2\mathcal{V}(\psi)&=&-3 \alpha  \Lambda-\frac{9 p \mathcal{\tilde{C}}_0^2 \Lambda }{2 \beta } +\frac{27 \mathcal{\tilde{C}}_0 ^3 \psi_1 e^{\frac{2 \beta  \psi }{\mathcal{\tilde{C}}_0 }} (3 \mathcal{\tilde{C}}_0  \Lambda  \psi_1+2 \beta  M)}{2 \beta }-\frac{3}{2} \mathcal{\tilde{C}}_0 ^2 e^{\frac{\beta  \psi }{\mathcal{\tilde{C}}_0 }}\sqrt[3]{\frac{3 \mathcal{\tilde{C}}_0  \psi_1-e^{-\frac{\beta  \psi }{\mathcal{\tilde{C}}_0 }}}{\beta }} \left(9 \mathcal{\tilde{C}}_0  \psi_1 e^{\frac{\beta  \psi }{\mathcal{\tilde{C}}_0 }}+1\right)\,.\nonumber \\
            &&\label{sol3aaa}
        \end{eqnarray}
    \end{itemize}
    Here,  $\psi_1$ is an integration constant. All these solutions are scalar fields on S-(A)dS spacetimes, solving the scalar-torsion field equations.  As mentioned before, instead of choosing the coupling function $\mathcal{\tilde{C}}(\psi)$ one can also choose either $\psi$ or $\mathcal{V}(\psi)$ and obtain new solutions. 
    
    Even for $\Lambda \to 0$, i.e.\ when the metric just becomes the Schwarzschild metric, we see that these theories allow for non-trivial scalar field solutions. However, these do not have any influence on the geometry since the integrand in the expression for the energy of the scalar field \eqref{energy} is zero. This is expected since we know from \eqref{eq:symGR} that the effective energy density of the scalar field must be proportional to $\Theta^{(\psi)}{}^{t}{}_{t}(r)$ which is equated to the $tt$ component of the Einstein tensor. The latter vanishes for a S-(A)dS spacetime.
    
    \item Second we consider $C\neq 1$ and a vanishing potential ($\mathcal{V}(\psi)=0$).
    
   Following a similar approach as before, it is possible to manipulate the field equations to get an equation that does not depend on the coupling function. To do this, we first solve~\eqref{eq:rrF2} for $\mathcal{\tilde{C}}(\psi)$ and \eqref{eq:ththF2} for $\mathcal{V}(\psi)$ and replace those expressions in the modified Klein-Gordon equation~\eqref{eq:psiF2}. Then, we assume that $\mathcal{V}(\psi)=0$ and rearrange the equation yielding
    \begin{eqnarray}\label{eqfea}
    0&=&r^2 A A'^2 \left(A^2 \left(6  -2 \beta  r^3 \psi' \psi''-\beta  r^2 \psi'^2\right)+10   C^2\right)+r A'\Big(A^2 C^2 \left(4  +\beta  r^2 \psi'^2\right)\nonumber\\
    &&+A^4 \left(6  -2 \beta  r^3 \psi' \psi''-\beta  r^2 \psi'^2\right)-2   C^4\Big)-2 r^3 A^2 A'^3 \left(6  +\beta  r^2 \psi'^2\right)+A'' \Big[2 r^3 A^3 A'\left(6  +\beta  r^2 \psi'^2\right)\nonumber\\
    &&+r^2 A^2 \left(A^2 \left(6  +\beta  r^2 \psi'^2\right)-6   C^2\right)\Big]+2 A^3 C^2 \left(2  +\beta  r^2 \psi'^2\right)-4   AC^4+4 \beta  r^2 A^5 \psi' \left(r \psi''+\psi'\right)\,.
    \end{eqnarray}
    To find solutions to this equations, we will further assume that the theory is $\mathcal{\tilde{C}}(\psi)=\mathcal{\tilde{C}}_0\psi$ and also that the scalar field behaves as $\psi=\psi_0 \log(r)$. By replacing these assumptions in~\eqref{eq:rrF2} and in the above equation we find 
    \begin{eqnarray}
    C(r)^2&=&\frac{A }{2  }\left(2 r A' (2  -\mathcal{\tilde{C}}_0 \psi_0 )+A (2  -\psi_0  (\beta  \psi_0 +4 \mathcal{\tilde{C}}_0))\right)\,,\\
    A(r)&=&r^{-\frac{\beta  \psi_0}{\mathcal{\tilde{C}}_0}-2} \left(A_0+r^{\frac{4   \beta  \psi_0+6   \mathcal{\tilde{C}}_0-\beta  \mathcal{\tilde{C}}_0 \psi_0^2}{2   \beta  \psi_0+6   \mathcal{\tilde{C}}_0}}\right)^{\frac{  \beta  \psi_0+3   \mathcal{\tilde{C}}_0}{2   \mathcal{\tilde{C}}_0-\mathcal{\tilde{C}}_0^2 \psi_0}}\,,
    \end{eqnarray}
    with $A_0$ being an integration constant. Finally, by replacing all these expressions in the remaining field equation we find that $\mathcal{\tilde{C}}_0=(-2  \pm \sqrt{4 -2  \beta  \psi_0^2})/\psi_0$. Thus, the final solution gives us the following form of the metric
    \begin{eqnarray}\label{solmetric2}
    \dd s^2 =r^{-2}\Big(A_0r^{\pm \sqrt{2w}}+r^{\pm\sqrt{w/2}}\Big)\dd t^2-\frac{A_0\sqrt{ w}}{2}r^{\pm\sqrt{w/2}}\Big( A_0 r^{\pm \sqrt{w/2}}+1\Big)^{-1}(\pm 2 \sqrt{2} -\sqrt{w})\dd r^2-r^2\dd \Omega^2\,,
    \end{eqnarray}
    with $w=2-\beta \psi_0^2$, corresponding to the coupling function, potential and scalar field being equal to
    \begin{eqnarray}\label{sol3complex}
    \mathcal{\tilde{C}}(\psi)=\Big(\frac{-2  \pm \sqrt{2w}}{\psi_0}\Big)\psi\,,\quad \mathcal{V}(\psi)=0\,,\quad \psi=\psi_0 \log(r)\,.
    \end{eqnarray}
    The metric~\eqref{solmetric2} does not obey $g_{tt}=-1/g_{rr}$ and it is always non-asymptotically flat. Note that $w\geq 0$ to ensure that the physical quantities are real.
\end{itemize}

Thus we demonstrated explicitly the existence of non-trivial scalar fields in scalar-torsion theories of gravity based on a non-minimal coupling to the teleparallel boundary term $B$, for the real and the complex tetrad in spherical symmetry. For a non-trivial scalar field all the solutions we found in this section were not asymptotically flat. In Sec.~\ref{sec:nohair}, we find no hair theorems for certain classes of scalar-torsion theories with non-minimal coupling to $B$, which help to narrow down the class of theories in which one should search for asymptotically flat scalarized solutions.

Next, we will see that further interesting solutions exist for a non-minimal coupling to the torsion scalar.

\subsection{\texorpdfstring{$\mathcal{\tilde{C}}(\psi)=0$\,:}\, only non-minimal coupling between the scalar field and the torsion scalar}\label{ssec:torsionscalarcoupling}

Another scalar tensor theory in teleparallel gravity is to introduce a non-minimal coupling between the torsion scalar and the scalar field, for which boson stars have been found \cite{Horvat:2014xwa}. Here we present several scalarized solutions for such theories for the real and the complex tetrad. In contrast to the previous section, in this section we find asymptotically flat scalarized black holes.

\subsubsection{Analytical analysis of the field equation for the real tetrad}\label{sssec:real2}
We follow a similar strategy as the previous section. By solving the equations~\eqref{eq:ttF}-\eqref{eq:ththF} for $\mathcal{A}(\psi), \mathcal{A}'(\psi)$ and $\mathcal{V}(\psi)$ and then replacing these expressions (and their derivatives) into~\eqref{eq:psiF} one finds a cumbersome equation which is generically independent of the potential $\mathcal{V}(\psi)$ and the coupling function $\mathcal{A}(\psi)$ and thus only depends on $A(r),\, C(r),\, \psi(r)$. In this case the equation is much more involved than~\eqref{eqGen1}. For completeness, this equation is displayed in the appendix (see Eq.~\eqref{eqappendix}).

Again, by making some assumptions on the solution we are looking for, we can solve the equation. Imposing that $C=1$ we can discuss two classes of solutions.

\begin{itemize}[leftmargin=*]
\item The first one is given by  the ansatz $A(r) = 1-\frac{K}{r^p}$ which implies the following metric
\begin{eqnarray}
 \dd s^2=\Big(1-\frac{K}{r^p}\Big)^2\dd t^2-\Big(1-\frac{K}{r^p}\Big)^{-2}\dd r^2-r^2\dd \Omega^2\,. \label{metric1}
\end{eqnarray}
By replacing this ansatz in the necessary constraint equation~\eqref{eqappendix} that does not depend on the coupling function and the potential, we find that $\psi$ is represented by Appell hypergeometric functions for any power-law $p$. In particular, three particular special cases can have less involved form for the scalar field, namely
\begin{eqnarray}
 \psi(r)&=&-\frac{2 \psi_0 \sqrt{r}}{K \sqrt{r-K}}\,,\quad p=1\,, \label{sol3a}\\
\psi(r)&=&\frac{2 \psi_0}{5 K \left(1-K r^2\right)^{5/4}}\,,\quad p=-2\,,\label{sol3b}\\
 \psi(r)&=&\frac{4 \psi_0\sqrt[4]{r} \left(1-\frac{\sqrt{r}}{K}\right)^{5/6} \, _2F_1\left(\frac{1}{2},\frac{11}{6};\frac{3}{2};\frac{\sqrt{r}}{K}\right)}{K \left(K-\sqrt{r}\right)^{5/6}}\,,\quad p=1/2\,, \label{sol3c}
\end{eqnarray}
where $_2F_1$ is the hypergeometric function. Interestingly, the first solution, with $p=1$, reproduces the metric called the Bocharova–Bronnikov–Melnikov–Bekenstein (BBMB) solution found in Riemannian conformal scalar-vacuum theory~\cite{Bekenstein:1975ts,Bekenstein:1974sf,bocharova1970exact}. This solution behaves as a extremal Reissner-Nordström black hole for $K\neq 0$ and reduces to the Minkowski metric for $K=0$. Clearly, the scalar field diverges at the horizon $r = r_h=K$. However, the energy of the scalar field~\eqref{energy} for this solution is $\mathcal{E}_\psi=4\pi K$ which is always positive and  finite. This solution was shown to be unstable against linear perturbations~\cite{Bronnikov:1978mx} but those studies were performed by considering the Riemannian conformal scalar-vacuum theory. In our case, this solution could have a different behaviour since the field equations are different. There are other claims about the BBMB solution related to the energy-momentum tensor which emphasises that it is ill-defined at the horizon~\cite{Sudarsky:1997te}.

For the cases $p=1$ and $p=-2$, we can solve the remaining field equations to get the following exact solutions 
\begin{eqnarray}
\mathcal{A}(\psi)&=&-\frac{1}{8}\beta \psi^2\,,\quad \mathcal{V}(\psi)=0\,,\quad p=1\,,\label{solV1}\\
\mathcal{A}(\psi)&=&\frac{5\beta}{8} \psi^2\,,\quad \kappa^2\mathcal{V}(\psi)=-\frac{15}{4}K\beta\psi^2 \,,\quad p=-2\,.\label{solV2}
\end{eqnarray}
Even though the case $p=1/2$ is an analytical solution to the field equations, it is not easy to invert $\psi=\psi(r)$ to $r=r(\psi)$ to then find an explicit form of the coupling function and the potential. 

\item The second class of exact solutions can be obtained by adding the assumption $\mathcal{V}(\psi)=0$ to $C=1$. Assuming furthermore that $r A'+A-1\neq 0$ (which would reduce to the solution~\eqref{metric1} with $p=1$ in case of an equality) and $2 r A A'+A^2-1\neq0$ (which would require $\psi=\textrm{const.}$ in case of an equality), one can solve~\eqref{eq:rrF} and \eqref{eq:ththF} for $\mathcal{A},\mathcal{A}'$ and then replace those expressions in~\eqref{eq:ttF} to find the following differential equation
\begin{eqnarray}
0&=&r A (A-1) A''+(A-1)^2 A'-r (A+1) A'^2\,,
\end{eqnarray}
which has the following solution for the metric
\begin{eqnarray}
\dd s^2=\Big(2-\frac{r}{2 K}+\frac{\sqrt{r (r-4 K)}}{2 K}\Big)^2\dd t^2-\Big(2-\frac{r}{2 K}+\frac{\sqrt{r (r-4 K)}}{2 K}\Big)^{-2}\dd r^2-r^2\dd \Omega^2\,,\label{solution5}
\label{metricB}
\end{eqnarray}
where $K$ is an integration constant. By replacing the above equation in the modified Klein-Gordon equation~\eqref{eq:psiF} we find that the scalar field is
\begin{eqnarray}
\psi(r)=\frac{\psi_0 \left(\sqrt{r (r-4 K)}-4 K+r\right) \sqrt[4]{\sqrt{r (r-4 K)}-2 K+r}}{3 K r^{3/4} \sqrt{r-4 K}}\label{sol4}\,.
\end{eqnarray}
Finally, we find that the coupling function must be of the form
\begin{eqnarray}\label{Asol}
\mathcal{A}(\psi)=\frac{3\beta}{8}\psi^2\,.
\end{eqnarray}
The solution~\eqref{metricB} is well defined for $K\to0$ and has one horizon at $r_h=4K$. Moreover it is asymptotically flat. The energy of the scalar field~\eqref{energy} is $\mathcal{E}_\psi=-8\pi K$. For $K>0$ there exists a well defined horizon, but the energy of the scalar field is negative, while for $K<0$ the energy of the scalar field is positive, but the spacetime does not possess a horizon, since there does not exist any $r>0$ such that $g_{tt}(r)=0$.

A series expansion for $K\ll 1$ gives us the following metric and scalar field up to second order in $K$ 
\begin{eqnarray}
\dd s^2&=&\Big(1-\frac{2 K}{r}-\frac{3 K^2}{r^2}\Big)\dd t^2-\Big(1-\frac{2 K}{r}-\frac{3 K^2}{r^2}\Big)^{-1}\dd r^2-r^2\dd \Omega^2+\mathcal{O}(K^3)\,,\\
\psi(r)&=&\frac{2 \sqrt[4]{2} \psi_0}{3 K}-\frac{\sqrt[4]{2} \psi_0}{r}-\frac{3 K \psi_0}{2\ 
\sqrt[4]{8} r^2}-\frac{35 K^2 \psi_0}{12 \sqrt[4]{8} r^3}+\mathcal{O}(K^3)\,,
\end{eqnarray}
which behaves similarly as a Reisser-Nordstrdöm (RN) metric with $K=M$ acting as a mass but the charge appearing with the opposite sign behaving as an imaginary charge ($Q^2=-3M^2$).
\end{itemize}

To summarize this section: we demonstrated that there exist several non-trivial scalarized solutions to scalar-torsion gravity with a non-minimal coupling to the torsion scalar, which goes with a coupling function $\mathcal{A}\sim \psi^2$ and $\mathcal{\tilde C} = 0$ for the real tetrad. In particular we also found asymptotically flat solutions. As we will see in Sec.~\ref{sec:nohair}, the classes of theories we considered here consistently evade the no-hair theorems since the potentials do not satisfy a condition which is necessary for the no-hair theorems to hold.

\subsubsection{Analytical analysis of the field equation for the complex tetrad}\label{sssec:complex2}
We will now find exact solutions for the complex tetrad~\eqref{tetrad2} for the class of theories defined by $\mathcal{\tilde{C}}(\psi)=0$.

Due to the structure of the field equations~\eqref{eq:ttF2}-\eqref{eq:psiF2}, it is convenient to separate the study into the case with vanishing and non-vanishing potential. The latter case turns out to be very complicated and it is not possible to find exact solutions, which is why we assume $\mathcal{V}(\psi)=0$ throughout this section.

When we solve~\eqref{eq:ttF2} and \eqref{eq:rrF2} for $\mathcal{A}'(\psi)$ and $\mathcal{A}(\psi)$ and replace these expressions into~\eqref{eq:ththF2}, we find, as in the previous cases, a model independent necessary equation that the metric components have to satisfy. For any non-trivial coupling function $\mathcal{A}(\psi)$ the following must hold
\begin{eqnarray}
C(r)^2=A \left[r A'-A \left(\frac{r A''}{A'}+1\right)\right]\,.\label{eq111}
\end{eqnarray}
\begin{itemize}[leftmargin=*]
    \item Let us again first consider the case $C(r)=1$. Then, it is straightforward to solve~\eqref{eq111} and then the remaining equations for the system~\eqref{eq:ttF2}-\eqref{eq:psiF2} can be manipulated in the following way. First, we can solve the equation~\eqref{eq:rrF2} for the coupling function and replace this expression into~\eqref{eq:ttF2}. The corresponding equation can be easily solved for $\psi$. Lastly, we invert $\psi=\psi(r)$ to $r=r(\psi)$ and solve the remaining equation for the coupling function. After doing all of this procedure, we find the following solution
    \begin{eqnarray}\label{solC1}
    \dd s^2&=&\Big(\frac{1}{2 p+1}-2 M r^{-2 p-1}\Big)\dd t^2-\Big(\frac{1}{2 p+1}-2 M r^{-2 p-1}\Big)^{-1}\dd r^2-r^2\dd \Omega^2\,,\\
    \mathcal{A}(\psi)&=&-\frac{1}{4}p\beta\,\psi^2\,,\quad \psi(r)=\psi_0 r^{p}\,,\quad \mathcal{V}(\psi)=0\,,\quad p\neq 0\,,
    \end{eqnarray}
    with $p,\,\psi_0$ and $M$ being integration constants. In this solution, we have assumed that $p\neq0$ and thus $\psi=\psi(r)$.  Again this solution is non-asymptotically flat.
        
    \item As second case we assume $C\neq1$. We can solve~\eqref{eq:ttF2} and \eqref{eq:rrF2} for $\mathcal{\tilde{C}}'(\psi)$ and $\mathcal{\tilde{C}}''(\psi)$ and replace them into  into the modified Klein-Gordon equation~\eqref{eq:psiF2}. If one further replaces~\eqref{eq111} and assumes $\mathcal{V}(\psi)=0$, one finds that the scalar field $\psi(r)$ and the metric function $A(r)$ must satisfy the following differential equation
    \begin{eqnarray}\label{eqa}
        0&=&5 r^3 A^3 A''^3 \psi'-3 r^3 A'^6 \psi'-r^2 A A'^5 \left(4 r \psi''+5 \psi'\right)+A'^4 \left(2 r^3 A A'' \psi'-4 r A^2 \left(r \psi''+2 \psi'\right)\right)\nonumber\\
        &&+A'^3 \left(r^3 A^2 A''' \psi'-2 r^2 A^2 A'' \psi'+8 A^3 \left(r \psi''+2 \psi'\right)\right)+A'^2 \Big(-4 r^2 A^3 A''' \psi'-2 r^3 A^2 A''^2 \psi'\nonumber\\
        &&+12 r A^3 A'' \left(r \psi''+2 \psi'\right)\Big)+A' \left(r^2 A^3 A''^2 \left(4 r \psi''+15 \psi'\right)-3 r^3 A^3 A''' A'' \psi'\right)\,.
    \end{eqnarray}
    To be able to solve this equation we consider two sub cases:
   \begin{itemize}[leftmargin=*]    
        \item A way to find a solution for~\eqref{eqa} is by imposing that the terms multiplying $A'''$ in~\eqref{eqa} vanish. After doing this, we can find the form of the scalar field by solving the equation~\eqref{eqa}. Then, to find a solution we just need to use the remaining field equation to get the coupling function. All this procedure gives the following exact solution
        \begin{eqnarray}
            \dd s^2&=&\frac{1}{r}\left(2+K r^{1/3}\right)^{3}\dd t^2-\frac{1}{3}\Big(1+\frac{2}{Kr^{1/3}}\Big)^{-1} \dd r^2-r^2\dd \Omega^2\,,\\
            \mathcal{A}(\psi)&=&\frac{3 \beta  e^{-\frac{\psi }{3}} \left(K e^{\psi /3}-1\right)^2}{4 K}\,,\quad \psi(r)=\log \left(\frac{r}{\left(K r^{1/3}+2\right)^3}\right)\,, \quad \mathcal{V}(\psi)=0\,.
        \end{eqnarray}
       This solution is non-asymptotically flat. 
    
        \item We now fix the coupling constant $\mathcal{A}(\psi)$ and then solve the field equations. Choosing $\mathcal{A}(\psi)=\mathcal{A}_0 \psi^2$ and also by imposing $\psi(r)=\psi_0 r^{p}$, we can easily solve the system~\eqref{eq:ttF2}-\eqref{eq:psiF2} which gives a more general solution than~\eqref{solC1}. This solution has the same form as~\eqref{solC1} for the scalar field $\psi(r)$ and the coupling function $\mathcal{A}(\psi)=-\frac{1}{4}p\beta  \psi^2$, meaning that the field equations are solved only when the coupling constant is $\mathcal{A}_0 = -\frac{p \beta}{4}$. We consider the non-trivial case $p\neq 0$. For this solution the metric does not obey $C\neq1$. Thus, the solution has the following form
        \begin{eqnarray}\label{eq:solk}
            \dd s^2&=&\Big(A_0-2 M r^{-2 p-1}\Big)\dd t^2-\Big(\frac{A_0(2  p+1)}{A_0-2 M r^{-2 p-1}}\Big)\dd r^2-r^2\dd \Omega^2\,.
        \end{eqnarray}
        Here, $M,\,p$ and $A_0$ are integration constants. Again, this solution is non-asymptotically flat. By choosing $A_0=\frac{1}{1+2p}$, one recovers the metric~\eqref{solC1}. It should be noted that the solution~\eqref{solC1} was found by only assuming $C(r)=1$ while the above solution was obtained by assuming the coupling function and also the form of the scalar field. This means that the solution~\eqref{solC1} is the most general one satisfying $C(r)=1$ for our theory while the above solution might not be the most general solution for the squared coupling theory. Observe that the metric \eqref{eq:solk} is a generalisation of the metric~\eqref{solC1} which we found earlier.
        
        Let us now study \eqref{eq:solk} by supposing that $p\ll 1$ and expanding the terms up its first order corrections. By doing this, we find that the metric and the scalar field quantities becomes Schwarzschild plus a modified correction related to the scalar field:
        \begin{align}\label{expanded1}
            \dd s^2&=\Big[1-\frac{2M}{r}+\frac{4M p}{r}\log(r)\Big]\dd t^2-\Big(1-\frac{2M}{r}\Big)^{-1}\Big[1+2 p-\Big(1-\frac{2M}{r}\Big)^{-1}\frac{4 M p }{r }\log (r)\Big]\dd r^2-r^2\dd \Omega^2+\mathcal{O}(p^2)\,,\\
            \mathcal{A}(\psi)&=-\frac{1}{4}p\beta \psi^2\,,\quad \psi(r)=\psi_0+p\phi_0 \log(r)+\mathcal{O}(p^2)\,,\quad  \mathcal{V}(\psi)=0\,, \quad p\neq 0\,.\label{expanded2}
        \end{align}
        Here, we have also set $A_0=1$ to recover Schwarzschild at the background level. Notice that since we assumed that, the above expanded metric would have a different behaviour as~\eqref{solC1}.
        This solution then corresponds to a scalarised black hole solution with $p$ being the parameter that controls the scalar field. It is worth mentioning that this expanded form of the metric contains a logarithmic term similarly as it was previously obtained in $f(T)$ and $f(T,B)$ gravity where perturbed solutions around Schwarzschild were found~\cite{Bahamonde:2019zea,Bahamonde:2020bbc,Pfeifer:2021njm,Bahamonde:2021srr,Bahamonde:2020vpb}.
    \end{itemize}
\end{itemize}

Thus, also for the complex tetrad we could find scalarized solutions for torsion scalar non-minimally coupled scalar fields, however none of them is asymptotically flat.

\section{On no-hair theorems in scalar-torsion theories of gravity}\label{sec:nohair}
In the context of scalar-tensor theories of gravity several no-hair theorems have been proven, which state that, under certain assumptions, there exist no non-trivial scalar fields on black hole spacetimes~\cite{Herdeiro:2015waa}.

In the previous section we demonstrated the existence of static spherically symmetric solutions in scalar-torsion theories for certain types of non-minimal coupling between the scalar field and the torsion of the teleparallel connection. Hence, a general no-hair theorem cannot be expected. 

However, when we study the field equations carefully we can derive some necessary constraints that have to be satisifed by the coupling functions $\mathcal{A}$ and $\tilde{\mathcal{C}}$ as well as by the potential $\mathcal{V}$ so that non-trivial scalar fields can exist.

On the one hand, taking the trace of the symmetric vacuum field equations \eqref{eqn:clafeqtets}, we find
\begin{align}
     - \mathcal{A} \mathring{R} -2 (\mathcal{A}' + \tilde{\mathcal{C}}')T^\mu\partial_\mu\psi + (\beta - 3 \tilde{\mathcal{C}}'')\mathring{\nabla}_\mu\psi\mathring{\nabla}^\mu\psi - 3 \tilde{\mathcal{C}}'\mathring{\square}\psi + 4 \kappa^2 \mathcal{V} = 0\,.
\end{align}
A little further manipulation of the $\mathring{\square}\psi$ term yields the convenient form
\begin{align}
  - \mathcal{A} \mathring{R} -2 (\mathcal{A}' + \tilde{\mathcal{C}}')T^\mu\partial_\mu\psi + \beta \mathring{\nabla}_\mu\psi\mathring{\nabla}^\mu\psi + 4 \kappa^2 \mathcal{V} = 3 \mathring{\nabla}_\mu (\tilde{\mathcal{C}}' \mathring{\nabla}_\mu\psi )\,.  \label{eq:traceTetradNH}
\end{align}

On the other hand the vacuum scalar field equation \eqref{eqn:clafeqscal} has to hold
\begin{align}
    \frac{1}{2}\mathcal{A}'(\psi)T+ \frac{1}{2}\mathcal{\tilde{C}}'(\psi)B  - \beta\mathring{\square}\psi+ \kappa^2\mathcal{V}' = 0\,,
\end{align}
which can be multiplied with $\psi$ to obtain
\begin{align}
    \mathring{\nabla}_\mu\left( \left(\mathcal{A}'+\mathcal{\tilde C}'\right)\psi T^\mu - \beta \psi \mathring{\nabla}^\mu \psi\right) + \beta \mathring{\nabla}_\mu\psi \mathring{\nabla}^\mu \psi - \frac{1}{2}\psi \mathcal{A}'\mathring{R} - \left(\mathcal{A}'+\mathcal{\tilde C}' + \psi \mathcal{A}'' + \psi \mathcal{\tilde C}'' \right)T^\mu\partial_\mu \psi + \kappa^2 \psi \mathcal{V}' = 0\,. \label{eq:scalarNH}
\end{align}

We can integrate equation \eqref{eq:traceTetradNH} and \eqref{eq:scalarNH} over any volume $V\subset M$ to obtain the constraints
\begin{align}
   \int_V \dd^4x\ \theta \left( \beta \mathring{\nabla}_\mu\psi\mathring{\nabla}^\mu\psi - \mathcal{A} \mathring{R} - 2 (\mathcal{A}' + \tilde{\mathcal{C}}')T^\mu\partial_\mu\psi  + 4 \kappa^2 \mathcal{V} \right)
   &= \int_{\partial V} 3 (\tilde{\mathcal{C}}' \mathring{\nabla}^\mu\psi )n_\mu \dd\sigma \label{eq:NH1}\\
   \int_V \dd^4x\ \theta \left( \beta \mathring{\nabla}_\mu\psi \mathring{\nabla}^\mu \psi - \frac{1}{2}\psi \mathcal{A}'\mathring{R} - (\mathcal{A}'+\mathcal{\tilde C}' + \psi \mathcal{A}'' + \psi \mathcal{\tilde C}'' )T^\mu\partial_\mu \psi + \kappa^2 \psi \mathcal{V}'\right)
   &= -\int_{\partial V} ( (\mathcal{A}'+\mathcal{\tilde C}')\psi T^\mu - \beta \psi \mathring{\nabla}^\mu \psi )n_\mu \dd\sigma\,,\label{eq:NH2}
\end{align}
where $n^\mu$ is the unit normal of the boundary of the volume $V$ and $\dd\sigma = \sqrt{h}\, \dd^3u$ is the pull-back of the volume element $\dd^4x\ \theta$ to $\partial V$ equipped with coordinates $u$.

From now on, we will assume that $M$ is a spherically symmetric asymptotically flat black hole spacetime and $V$ its exterior region. The boundary of this volume $\partial V$ is thus given by the event horizon $\mathcal{H}$ and the asymptotic flat region $\mathcal{H}_\infty$ ($A(r)\to 1, C(r)\to 1$) at infinity. Regarding the behaviour of the scalar field, we will assume that $\psi$ only depends on $r$ (to respect the spherical symmetry) and that approaching $\mathcal{H}_\infty$ we have $\partial_\mu \psi \to 0$. 

We know that, at the horizon $\mathcal{H}$, the non-vanishing components of the normal are given by $n^t$ only (since $\mathcal{H}$ is a Killing horizon). In addition, for the spherically symmetric tetrads which we employ, we found that the vector torsion has only a non-vanishing $T^r$ component, see \eqref{Tmu1} and \eqref{Tmu2}. Consequently, the boundary term in \eqref{eq:NH1} vanishes, while the boundary term in \eqref{eq:NH2} reduces to
\begin{align}
    \int_{\mathcal{H}_\infty} \left(\left(\mathcal{A}'+\mathcal{\tilde C}'\right)\psi T^\mu\right)n_\mu \dd\sigma\,.
\end{align}
This boundary term vanishes for both the real and complex tetrad, since in both cases, the vector torsion vanishes where $A(r)\to1, A'(r)\to 0$ and $r\to\infty$ (in the asymptotic flat regions), see again~\eqref{Tmu1} and~\eqref{Tmu2}.

Let us continue with the analysis of the equations \eqref{eq:NH1}  and \eqref{eq:NH2}. Under the above assumptions, these equations reduce to
\begin{align}
   \int_V \dd^4x\ \theta \left( \beta \mathring{\nabla}_\mu\psi\mathring{\nabla}^\mu\psi - \mathcal{A} \mathring{R} - 2 \left(\mathcal{A}' + \tilde{\mathcal{C}}'\right)T^\mu\partial_\mu\psi  + 4 \kappa^2 \mathcal{V} \right)
   &= 0 \,, \label{eq:NH11}\\
   \int_V \dd^4x\ \theta \left( \beta \mathring{\nabla}_\mu\psi \mathring{\nabla}^\mu \psi - \frac{1}{2}\psi \mathcal{A}'\mathring{R} - \left(\mathcal{A}'+\mathcal{\tilde C}' + \psi \mathcal{A}'' + \psi \mathcal{\tilde C}'' \right)T^\mu\partial_\mu \psi + \kappa^2 \psi \mathcal{V}'\right)
   &= 0\,.\label{eq:NH22}
\end{align}
Substracting the two equations yields
\begin{align}
   \int_V \dd^4x\ \theta \left(  \left(\mathcal{A} - \tfrac{1}{2}\psi \mathcal{A}'\right)\mathring{R} + \left( \mathcal{A}' + \mathcal{\tilde C}' - \psi \mathcal{A}'' - \psi \mathcal{\tilde C}'' \right)T^\mu\partial_\mu \psi+ \kappa^2 \left(\psi \mathcal{V}' - 4 \mathcal{V}\right) \right) = 0\,.\label{eq:NH4}
\end{align}
Thus, in total we find the necessary constraints \eqref{eq:NH11}, \eqref{eq:NH22} and \eqref{eq:NH4}, which need to be satisfied in order for a non-trivial scalar field to exist. They impose restrictions on the kinetic term of the scalar field, the coupling functions and the potential. The third constraint~\eqref{eq:NH4} is of particular importance, since it allows us to replace the integral over the curvature scalar or the the vector torsion in \eqref{eq:NH11} and \eqref{eq:NH22} for several models. These then yield a constraint on the potential for the existence of non-trivial scalar fields in classes of scalar-torsion theories of gravity.

We explicitly discuss four cases:
\begin{enumerate}[leftmargin=*]
    \item A pure polynomial coupling to the torsion scalar by choosing $\mathcal{A} = \alpha \psi^m$ and $ \tilde{\mathcal{C}}=0$. In this case equation \eqref{eq:NH4} can be solved for the integral over the Ricci scalar
    \begin{align}
        - \int_V \dd^4x\ \theta\ \alpha \psi^m \mathring{R} = \int_V \dd^4x\ \theta \left( 2 m \alpha \psi^{m-1} T^\mu\partial_\mu \psi + \frac{2 \kappa^2}{2-m}(\psi \mathcal{V}'-4 \mathcal{V} ) \right).
    \end{align}
    Using this result in equations \eqref{eq:NH11} and \eqref{eq:NH22} gives from both equations the same constraint
    \begin{align}\label{eq:NHcase11}
       \int_V \dd^4x\ \theta\left(\beta \lc{\nabla}_\mu\psi \lc{\nabla}^\mu\psi + \frac{2\kappa^2}{m-2}\left( 2 m \mathcal{V} - \psi \mathcal{V'}\right)\right)=0\,.
    \end{align}
   Consequently, for all potentials $\mathcal{V}$ satisfying
    \begin{align}\label{eq:NHcase111}
        \frac{2}{\beta (m-2)}\left( 2 m \mathcal{V}(\psi) - \psi \mathcal{V'}(\psi)\right) \leq 0\,, \quad\forall \,\psi(r)\,,
    \end{align}
    non-trivial scalar field solutions cannot exist, since $\lc{\nabla}_\mu \psi \lc{\nabla}^\mu \psi  = - \frac{C^2}{A^2}  \psi'^2 \leq 0$. If the inequality \eqref{eq:NHcase111} holds strictly, there  cannot exist any scalar-field solution, since it would immediately contradict \eqref{eq:NHcase11}, while if the equality in \eqref{eq:NHcase111} holds, then \eqref{eq:NHcase11} implies that the scalar field must be constant.
    
    \item Choosing $\mathcal{A} = \alpha \psi^2$ and $\psi\, \tilde{\mathcal{C}}''= \tilde{\mathcal{C}}'$ (i.e.\ $\tilde{\mathcal{C}}=\tfrac{c_1}{2}\psi^2+c_2$, in particular also for $\tilde{\mathcal{C}} = 0$), gives from \eqref{eq:NH4} directly
    \begin{align}\label{eq:NHCase1}
            \int_V \dd^4x\ \theta\ \kappa^2 (\psi \mathcal{V}' - 4 \mathcal{V}) = 0\,.
    \end{align}
    Hence for potentials $\mathcal{V}$ for which either $\psi \mathcal{V}' > 4 \mathcal{V}$ or $\psi \mathcal{V}' < 4 \mathcal{V}$ ($\forall \,\psi$) holds, non-trivial scalar field cannot exist. The solutions we found in \eqref{solV1}, \eqref{solV2} and around \eqref{Asol} evade this constraint for the potential. In particular for polynomial potential $\mathcal{V} = \psi^n$ we find that $\int_V \kappa^2 (n - 4)\psi^n = 0$ which cannot be satisfied for a non-trivial scalar field and $n$ being even, except for $n=4$.
    
    \item Choosing $\mathcal{A} = \alpha$ and $\mathcal{\tilde C}' + \psi \mathcal{\tilde C}'' = 0$ (i.e.\ $\mathcal{\tilde C} = c_1 \ln(\psi) + c_2$), turns the constraint \eqref{eq:NH22} into 
    \begin{align}\label{eq:NHCase2}
        \int_V \dd^4x\ \theta \left( \beta \mathring{\nabla}_\mu\psi \mathring{\nabla}^\mu \psi + \kappa^2 \psi \mathcal V'\right) = 0\,.
    \end{align}
    Hence, for this logarithmic non-minimal coupling to the boundary term, $\psi = \psi(r)$, we obtain a class of scalar-torsion theories of gravity for which a scalar no-hair Theorem holds similar to the minimally coupled general relativity, namely, there cannot be a scalarized static spherically symmetric black hole for any potential $\mathcal{V}$ which satisfies $\frac{\psi \mathcal{V}'}{\beta}\leq0$.
    
    \item Choosing a polynomial coupling to the boundary term  $\tilde{\mathcal{C}}=\frac{\gamma}{m+1}\psi^{m+1}$ and $\mathcal{A} = \alpha$ gives from \eqref{eq:NH4}
    \begin{align}\label{eq:NH61}
      -\int_V \dd^4x\ \theta \ \alpha \mathring{R} =  \int_V \dd^4x\ \theta  \left( (1-m) \gamma \psi^{m}  T^\mu\partial_\mu \psi  + \kappa^2 (\psi \mathcal{V}' - 4 \mathcal{V})\right) \,,
    \end{align}
    Using this in equation \eqref{eq:NH11}, as well as evaluating \eqref{eq:NH22} which is already independent of the Ricci scalar, both gives
    \begin{align}\label{eq:NH62}
        \int_V \dd^4x\ \theta  \left(\beta \mathring{\nabla}_\mu\psi \mathring{\nabla}^\mu \psi - (m+1) \gamma \psi^m T^\mu\partial_\mu \psi  + \kappa^2 \psi \mathcal{V}'  \right) = 0\,.
    \end{align}
    From this derived constraint we find that there cannot exist non-trivial scalar field profile if the potential $\mathcal{V}$ under consideration satisfies either
    \begin{align}
        \frac{1}{\beta }\left(\psi \mathcal{V}' - (m+1) \gamma \psi^m T^r\psi'\right) \leq 0\,, \quad \forall \quad r>r_h\,,
    \end{align}
    or, using \eqref{eq:NH62} in \eqref{eq:NH61}
    \begin{align}
        \frac{(m+1)}{m-1} \frac{1}{\beta} \left( \alpha \lc{R}+\kappa^2(\psi \mathcal{V}' - 4\mathcal{V})\right)\leq 0\,, \quad \forall \quad r>r_h\,.
    \end{align}
    Since these inequalities depend on the spacetime torsion, resp. the spacetime curvature, we cannot formulate these conditions in terms of $\psi$ alone. The interplay between the potential and the properties of spacetime needs to be taken into account.
\end{enumerate}

Let us summarize our findings in a first no-hair theorem for scalar-torsion theories:

\begin{theorem}\label{thm:nohair}
    Consider a scalar-torsion theory of gravity defined by the action \eqref{C}, the tetrads \eqref{tetrad} or \eqref{tetrad2} and a scalar field $\psi=\psi(r)$. There exist no spherically symmetric asymptotically flat scalarized black holes for the following couplings and potentials satisfying the corresponding displayed inequalities:
    \begin{enumerate}
        \item $\mathcal{A} = \alpha \psi^m$, $\tilde{\mathcal{C}}=0$ and $\frac{2}{\beta (m-2)}\left( 2 m \mathcal{V} - \psi \mathcal{V}'\right) \leq 0$;
        \item $\mathcal{A} = \alpha \psi^2$, $\tilde{\mathcal{C}}=\frac{c_1}{2}\psi^2 + c_2$ and either $\psi \mathcal{V}' > 4 \mathcal{V}$ or $\psi \mathcal{V}' < 4 \mathcal{V}$;
        \item $\mathcal{A} = \alpha$, $\mathcal{\tilde C}= c_1 \ln(\psi) + c_2$ and $\frac{\psi \mathcal{V}'}{\beta}\leq0$;
        \item $\mathcal{A} = \alpha$, $\tilde{\mathcal{C}}=\frac{\gamma}{m+1}\psi^{m+1}$ and $\frac{1}{\beta }\left(\psi \mathcal{V}' - (m+1) \gamma \psi^m T^r\psi'\right) \leq 0$ or $\frac{(m+1)}{m-1} \frac{1}{\beta} \left( \alpha \lc{R}+\kappa^2(\psi \mathcal{V}' - 4\mathcal{V})\right)\leq 0$.
    \end{enumerate}
\end{theorem}
Observe that a vanishing potential $\mathcal{V}=0$ (and thus $\mathcal{V}'$=0) evades the no-hair constraints in Case 2. For the Cases 1 and 3, a vanishing potential implies that there are no asymptotically flat spherically symmetric scalarized black hole solutions.
Finally, in Case 4, the existence of scalarized black holes is connected to the sign of the vector torsion, or equivalently, the Ricci scalar.

No hair theorems of this type can easily be obtained for further scalar-torsion theories of gravity, i.e.\ different choices of $\mathcal{A}$ and $\tilde{\mathcal{C}}$ as we presented here, by employing the algorithm we outlined below \eqref{eq:NH4}.

\section{Conclusions}\label{sec:conclusion}
In this paper, we studied spherical symmetry in scalar-torsion theories of gravity by considering a scalar field non-minimally coupled to both the torsion scalar and the teleparallel boundary term. This theory contains different subclasses that have been studied in both teleparallel gravity and the standard Riemannian scalar-tensor gravity. For instance, when the non-minimal coupling is reduced to the minimal case (setting $\mathcal{A}(\psi)=\mathcal{\tilde{C}}(\psi)=1/2$ and $\mathcal{B}(\psi)=\kappa^2$in~\eqref{C}), the theory obtained is the standard Riemannian scalar-tensor theory minimally coupled with the Ricci scalar with the Einstein-Klein-Gordon Lagrangian $\mathcal{L}=\frac{1}{4\kappa^2}\lc{R}-\frac{1}{2}(\partial_\mu \psi)(\partial^\mu \psi)-\mathcal{V}(\psi)$. Further, by choosing instead $\mathcal{\tilde{C}}(\psi)=- \mathcal{A}(\psi)$ in~\eqref{C}, the theory is extended to have a non-minimal coupling between the scalar field and the Riemannian Ricci scalar of the form $\mathcal{A}(\psi)\lc{R}$. In all the other non-trivial cases, the theory would have couplings that cannot be obtained in the Riemannian case and they can be considered to be related only to teleparallel gravity.

As usual in teleparallel theories of gravity the field equations can be decomposed into a symmetric~\eqref{eqn:clafeqtets} and antisymmetric part~\eqref{eq:asym}. By imposing spherical symmetry, one can solve the antisymmetric field equation in two different ways : one the one hand with a real tetard expressed in~\eqref{tetrad} and, on the other hand with a complex one given by~\eqref{tetrad2}. Consequently, the symmetric field equations in spherical symmetry have two branches which were presented in Secs.~\ref{sec:real}-\ref{sec:complex}. We studied these equations in two main teleparallel scalar-torsion theories, one with only a non-minimal coupling between the boundary term and the scalar field (see Sec.~\ref{ssec:scalarbdryy}) and another one with a only a non-minimal coupling between the torsion scalar and the scalar field (see Sec.~\ref{ssec:torsionscalarcoupling}). For these two theories, we split the study for the two possible tetrads and we presented exact solutions to scalar-torsion theories of gravity for the first time and thus demonstrated the existence of scalarized black holes in these theories. 

For the boundary term coupled theory (see Sec.~\ref{ssec:scalarbdryy}), we found two exact spherically symmetric solutions for the real tetrad, see Eqs.~\eqref{solmetric1}-\eqref{solpsi} and \eqref{solmetric2B}-\eqref{solpsi33}, respectively. These solutions are non-asymptotically flat and both metrics have one horizon. For the complex tetrad, we found S-(A)dS as exact solutions~\eqref{metricCase1} for different non-minimal couplings (see~\eqref{sol1b}-\eqref{sol3aaa}). One notices that even for the $\Lambda=0$ case, the scalar field can have a non-trivial profile leading to a Schwarzschild geometry endowed with a non-trivial scalar field. Moreover, for the complex tetrad we also found a non-asymptotically flat solution described by a non-trivial scalar field~\eqref{sol3complex} and a power-law type form of the metric~\eqref{solmetric2}.

In the theory which is defined by a coupling between the torsion scalar and the scalar field (see Sec.~\ref{ssec:torsionscalarcoupling}), we found three exact solutions for the real tetrad. The first one has the same metric as~\eqref{metric1} with $p=1$ which is the so-called BBMB black hole. The form of the scalar field and the coupling functions were displayed in~\eqref{sol3a} and~\eqref{solV1}. This solution is asymptotically flat and it has been found before in a Riemannian scalar-tensor theory which is conformally flat. Another solution that we found that leads to a non-asymptotically flat metric is described by~\eqref{metric1} with $p=-2$ and the scalar field, potential and couplings are given by~\eqref{sol3b} and \eqref{solV2}. The last solution for the real tetrad we found has an asymptotically flat metric~\eqref{solution5} with a non-trivial scalar field~\eqref{sol4} having a zero potential with a coupling of the form $\mathcal{A}=\frac{3\beta}{8}\psi^2$. When the constant related to the scalar field is assumed to be small, the metric becomes a RN-like metric with the charge being equal to the mass (extremal RN) but having an opposite sign in the charge-type term as in the RN metric. For the complex tetrad (see Sec.~\ref{sssec:complex2}), we obtained three exact solutions and all of them are non-asymptotically flat. For the solution~\eqref{eq:solk}, one notices that when the scalar field contribution becomes small, the metric~\eqref{expanded1} can be written approximately as a Schwarzschild modified metric with the scalar field acting as an extra term that modifies the spacetime to be non-asymptotically flat.

To summarize, we found several non-asymptotically flat solutions and, most noteworthy, two asymptotically flat scalarized solutions \eqref{metric1} and \eqref{metricB} which emerge for a non-minimal coupling between the scalar field $\psi$ and the torsion scalar $T$ which is proportional to $\psi^2$. 

As a natural complement to this analysis, we have also investigated no-scalar-hair arguments limiting the sectors in which spherically symmetric scalarized black holes can be found in these theories. These results are summarized in Theorem \ref{thm:nohair}.

\vspace{11pt}

This paper is a first step in the systematic analysis of the existence of hairy black-holes in scalar-torsion theories of gravity. It will be continued by extending the investigations on teleparallel Gauss-Bonnet scalar-torsion theory, such as the theories discussed in~\cite{Bahamonde:2016kba,Bahamonde:2020vfj,Bahamonde:2018ibz}. In the Riemannian case, it is well known that these theories have asymptotically flat scalarized black holes with spontaneous scalarization~\cite{Antoniou:2017acq,Silva:2017uqg,Brihaye:2015qtu}. Since the Gauss-Bonnet scalar-torsion theories contains the standard Riemannian case in a certain limit, it is obvious to mention that those theories also will contain those solutions. However, the nature of the pure teleparallel part is unknown and it would be interesting to explore those theories to find what kind of new scalarized black hole solutions can appear. Further the study of pseudo scalar/axion couplings \cite{Hohmann:2020dgy} will also be extended to spherical symmetry. 

Finally, a next step is to extend the results found here to rotating teleparallel black holes in axial symmetry \cite{Bahamonde:2020snl,Jarv:2019ctf}.

\begin{acknowledgments}
S.B. is supported by JSPS Postdoctoral Fellowships for Research in Japan and KAKENHI Grant-in-Aid for Scientific Research No. JP21F21789. S.B. also acknowledges the Estonian Research Council grants PRG356 ``Gauge Gravity"  and the European Regional Development Fund through the Center of Excellence TK133 ``The Dark Side of the Universe". C.P. was funded by the Deutsche Forschungsgemeinschaft (DFG, German Research Foundation) - Project Number 420243324.
\end{acknowledgments}

\appendix

\section{Non-minimally coupling between scalar field and torsion scalar - real tetrad}\label{app:Appx}
For the real tetrad, one can use the field equations~\eqref{eq:ttF}-\eqref{eq:psiF} to derive the following equation that does not depend on the form of the potential nor the coupling function:
\begin{eqnarray}
    0&=&\left(A^2+2 r A' A-C^2\right) \Big[\left(\left(3 C-r C'\right) \psi'+2 r C \psi''\right) A^5+A^4\Big\{C' \psi' \left(A'+r A''\right) r^2+C \Big(C' \psi'-A' \left(3 \psi'+2 r \psi''\right)\nonumber\\
    &&+r \left(r \psi' A'''-A'' \left(3 \psi'+2 r \psi''\right)\right)\Big) r-C^2 \left(5 \psi'+4 r \psi''\right)\Big\} +A^3\Big\{A' C' \psi' \left(A'+C'-r A''\right) r^3-C \Big(\left(\psi'+2 r \psi''\right) A'^2\nonumber\\
    &&+\left(C' \left(3 \psi'+2 r \psi''\right)+r \left(-2 r A'' \psi''-\psi' \left(2 A''+C''-r A'''\right)\right)\right) A'-2 r^2 \psi' A''^2\Big) r^2+C^2 \Big\{3 C' \psi'+A' \left(3 \psi'+2 r \psi''\right)\nonumber \\
    &&+r \left(A'' \left(3 \psi'+4 r \psi''\right)-2 r \psi' A'''\right)\Big\} r-2 C^3 \psi'\Big\} +A^2\Big\{C A' \Big(\psi' \left(-A'^2+\left(r \left(A''-C''\right)-2 C'\right) A'+C' \left(C'+3 r A''\right)\right)\nonumber \\
    &&+2 r A' \left(A'+C'\right) \psi''\Big) r^3+C^2r^2 \Big\{2 \left(\psi'+2 r \psi''\right) A'^2-\left(C' \left(3 \psi'-2 r \psi''\right)+r \left(2 r A'' \psi''+\psi' \left(3 A''+C''-r A'''\right)\right)\right) A'\nonumber\\
    &&-r \psi' A'' \left(C'+2 r A''\right)\Big\} +rC^3 \left(-5 C' \psi'+A' \left(\psi'+2 r \psi''\right)+r \left(A'' \left(3 \psi'-2 r \psi''\right)+r \psi' A'''\right)\right) -r^4 A'^2 C' \left(A'+C'\right) \psi'\nonumber\\
    &&+C^4 \left(6 \psi'+4 r \psi''\right)\Big\} +AC \Big\{\psi' \Big(2 A'^3 \left(A'+2 C'\right) r^4+C A'^2 \left(3 A'+C'-2 r A''\right) r^3+C^2 A' \left(3 A'+5 C'+r A''\right) r^2\nonumber\\
    &&+C^3 \left(A'+2 C'-3 r A''\right) r-C^4\Big)-2 r C \left(C+r A'\right) \left(C^2+r^2 A'^2\right) \psi''\Big\} -C^2 \left(C^2+r^2 A'^2\right) \left(C^2+2 r A' C+3 r^2 A'^2\right) \psi'\Big]\,.\nonumber \label{eqappendix} \\
\end{eqnarray}
\bibliographystyle{utphys}
\bibliography{TPSBH}

\end{document}